\documentclass[pre,aps,twocolumn,showpacs]{revtex4}
\usepackage{amsmath,amssymb,graphicx}
\usepackage{epsfig}
\usepackage{textcomp}

\begin{document}
\title{Extinction of metastable stochastic populations}
\author{Michael Assaf and Baruch Meerson} \affiliation{Racah Institute of Physics, Hebrew University of
Jerusalem, Jerusalem 91904, Israel}
%87.23.Cc 	Population dynamics and ecological pattern formation
%02.50.Ga 	Markov processes
\begin{abstract}

We investigate the  phenomenon of extinction of a long-lived self-regulating stochastic population, caused by intrinsic (demographic) noise. Extinction typically occurs via one of two scenarios depending on whether the absorbing state $n=0$ is a repelling (scenario A) or attracting (scenario B) point of the \textit{deterministic rate equation}. In scenario A the metastable stochastic population resides in the vicinity of an attracting fixed point next to the repelling point $n=0$. In scenario B there is an intermediate repelling point $n=n_1$ between the attracting point $n=0$ and another attracting point $n=n_2$ in the vicinity of which the metastable population resides. The crux of the theory is a dissipative variant of WKB (Wentzel-Kramers-Brillouin) approximation which assumes that the typical population size in the metastable state is large. Starting from the master equation, we calculate the quasi-stationary probability
distribution of the population sizes and the (exponentially long) mean time to extinction for each of the two scenarios.  When necessary, the WKB approximation is complemented (i) by a recursive solution of  the quasi-stationary master equation at small $n$ and (ii) by  the van Kampen system-size expansion, valid near the fixed points of the deterministic rate equation.
The theory yields both entropic barriers to extinction and pre-exponential factors, and holds for a general set of multi-step processes when detailed balance is broken. The results simplify considerably for single-step processes and near the characteristic
bifurcations of scenarios A and B.
\end{abstract}
\maketitle

\section{Introduction}
Extinction  of an isolated stochastic population
after maintaining a long-lived state  is a dramatic phenomenon.  It occurs,
even in the absence
of environmental variations, because of
an unusual chain of random events when population losses dominate over
gains.  Population extinction risk is a key
negative factor in viability of small populations
\cite{bartlett,assessment}, whereas extinction of a disease following an epidemic outburst
\cite{bartlett,epidemic} is of course favorable. The possibility and consequences of
extinction of biologically important components, regulated by
chemical reactions in living cells, have also attracted
interest \cite{bio}.  As stochastic population dynamics are usually far from
equilibrium, and no general methods of evaluating large fluctuations are available,
they are of much interest to physics
\cite{kampen,gardiner}.

This work deals with an isolated single-species
population undergoing a set of gain-loss processes.
We will assume that the population is well mixed, so that spatial degrees of freedom are irrelevant. At the level of the \textit{deterministic rate equation} (henceforth \textit{rate equation}),
which describes the time history of the mean population size $\bar{n}(t)$
and ignores fluctuations, $\bar{n}(t)$ flows to an
attracting fixed point, where the gain and loss processes balance
each other. The actual stochastic population, however, behaves
differently and ultimately becomes extinct. This is because, in the
absence of influx of new individuals,
the empty state $n=0$ is \textit{absorbing}: the probability of exiting from it is zero \cite{noexplosion}.

\begin{figure}
\includegraphics[width=3.18in, height=0.7in,clip=]{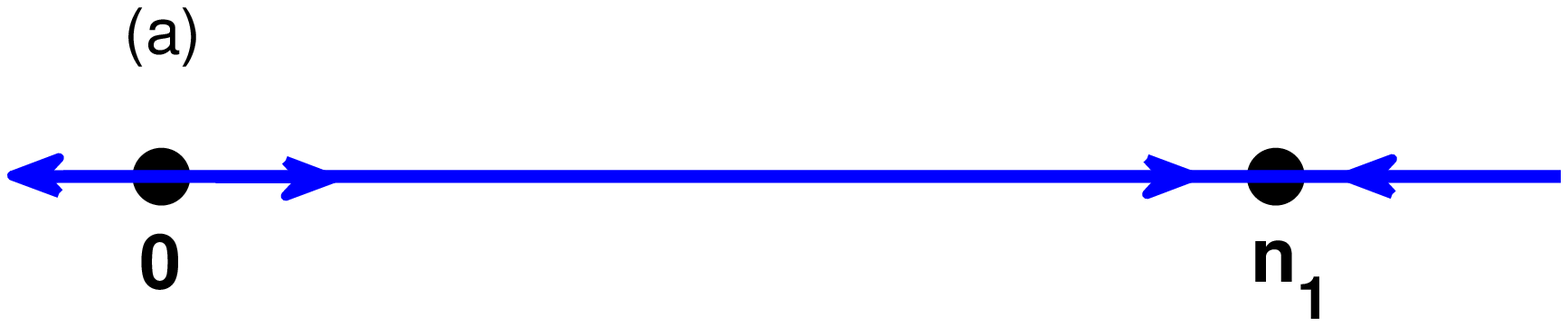}
\includegraphics[width=3.0in, height=0.7in,clip=]{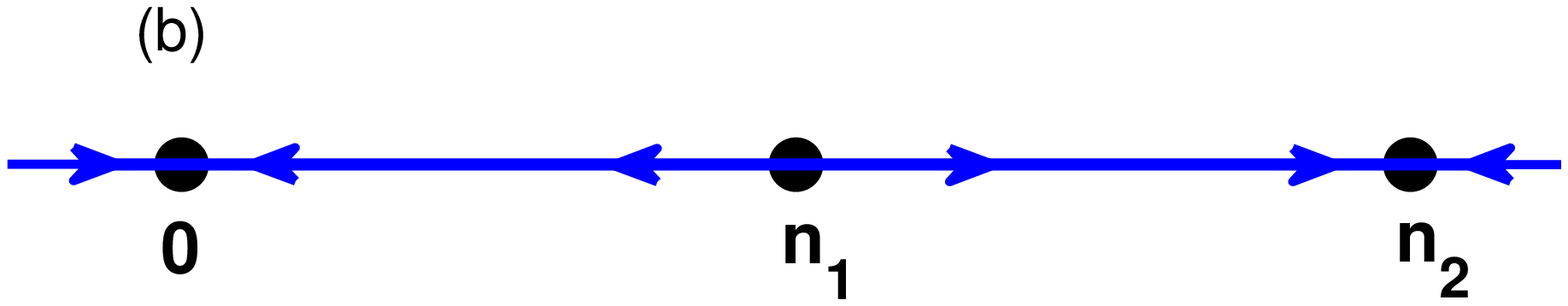}
\caption{(color online). Typical extinction scenarios are determined by the character of the fixed point $\bar{n}=0$ of the rate equation
(\ref{rateeqgen}). (a) Scenario A: the fixed point $\bar{n}=0$ is repelling. In the \textit{stochastic} system
extinction occurs via a large fluctuation which brings the metastable population
from a vicinity of the attracting fixed point $\bar{n}=n_1$ of the rate equation directly to the
absorbing state $n=0$. (b) Scenario B: the fixed point
$\bar{n}=0$ is attracting. In the stochastic system
extinction occurs via a large fluctuation which brings the metastable population
from a vicinity of the \textit{next} attracting fixed point $\bar{n}=n_2$ of the rate equation to a vicinity of the
repelling fixed point $\bar{n}=n_1$. From there the population flows ``downhill"
to the absorbing state $n=0$ almost deterministically.} \label{mf}
\end{figure}

Although extinction (and fluctuations in general) are beyond its scope, the rate equation is a convenient starting point of our analysis. For an isolated single-species population the rate equation can be written as
\begin{equation}\label{rateeqgen}
\frac{d\bar{n}}{dt}=\bar{n}\,\Phi(\bar{n})\,,
\end{equation}
where $\Phi(x)$ is a smooth function determined  by the specific gain-loss processes, see below.
For generic gain-loss processes $\Phi^{\prime}(0)\neq 0$. For $\Phi^{\prime}(0)>0$ the fixed point $\bar{n}=0$ is repelling, whereas for  $\Phi^{\prime}(0)<0$ it is attracting. In the former case, the next fixed point $\bar{n}=n_1>0$ of Eq.~(\ref{rateeqgen}) is attracting, see Fig.~\ref{mf}a. According to the rate equation, the mean population size in this case  flows to $\bar{n}=n_1$ and stays there forever. When varying the rate constants of the gain-loss processes, the attracting fixed point $\bar{n}=n_1$ emerges via a transcritical bifurcation.

Now let $\bar{n}=0$ be an attracting fixed point of the rate equation (\ref{rateeqgen}). To have a long-lived population of a nonzero size, at least two more fixed points of the rate equation (\ref{rateeqgen}) must be present: a repelling point $\bar{n}=n_1>0$ and an attracting point $\bar{n}=n_2>n_1$, see Fig.~\ref{mf}b. When starting from any $\bar{n}(t=0)>n_1$, the mean population size flows to $\bar{n}=n_2$  and, according to the rate equation, stays there forever.  The characteristic bifurcation in this case is saddle-node.

As we will see shortly, these two cases give rise to two different extinction scenarios of stochastic populations. To account for the intrinsic noise, we employ the master
equation
\begin{equation}\label{master}
\frac{d P_n(t)}{dt} = \sum_{r} \left[W_r(n- r)P_{n-r}(t) -W_r(n)P_n(t)\right]
\end{equation}
which describes the evolution of  the
probability $P_n(t)$ to have $n$ individuals at time $t$. Here $W_r(k)\geq 0$ is the transition rate between the states with $k$ and  $k+r$ individuals, whereas $r=\pm 1,\pm 2,\dots$, and all  terms that include
$P_{k}$ with $k<0$ are assumed to be zero. For $P_0(t)$ the master equation is
\begin{equation}\label{master0}
    \frac{d P_0(t)}{dt} = \sum_{r<0} W_r(-r)P_{-r}(t) \,.
\end{equation}
For $n=0$ to be an absorbing state, the process rates must obey, for any $r=\pm 1, \pm 2,\dots$, the condition $W_r(0)=0$.

We will be interested in
the important  regime of parameters for which the mean population size in the metastable state, as predicted by Eq.~(\ref{rateeqgen}), is large compared to one. Here,
prior to extinction, a long-lived probability distribution function (PDF) of the
population sets in, on a relaxation time scale $t_r$, around the corresponding attracting fixed point of the rate equation.  This long-lived PDF,
however, is metastable: it slowly decays
in time. Simultaneously, the probability to find the population extinct
slowly grows in time, see e.g. Refs.~\cite{Assaf,Assaf1}:
\begin{equation}\label{qsdintro}
P_{n>0}(t\gg t_r)\simeq \pi_n e^{-t/\tau}\;,\;\;P_0(t\gg t_r)\simeq
1-e^{-t/\tau}\,.
\end{equation}
The shape function $\pi_n$ ($n=1,2, \dots $) of the metastable PDF is called the
quasi-stationary distribution (QSD). For metastable populations a very strong inequality, $\tau \ggg t_r$ holds, and the
decay time $\tau$ is equal to the mean
time to extinction (MTE): the mean time it takes the stochastic
process to reach the absorbing state at $n=0$. The main objectives of this work is to accurately, and analytically, calculate the QSD $\pi_n$ and the MTE $\tau$ of a population which experiences quite a general set of stochastic gain-loss processes. The crux of the method is a dissipative WKB approximation \cite{bender,kubo,dykman}, where one assumes $n\gg1$, treats $n$ as a continuous variable and searches for $\pi_n$ as
\begin{equation}
\pi_n=e^{-NS(n)-S_1(n)-(1/N)S_2(n)-\dots}\,. \label{n30}
\end{equation}
Here $N \gg 1$ is a large parameter which scales as the mean population size in the metastable state. $S(n)$ is called the action, whereas $a(n)=e^{-S_1(n)}$ is called the amplitude.  The WKB approximation breaks down
at $n={\cal O}(1)$. Here a different approximation must be used, as explained below.

\begin{figure}
\includegraphics[width=2.6in,height=1.6in,clip=]{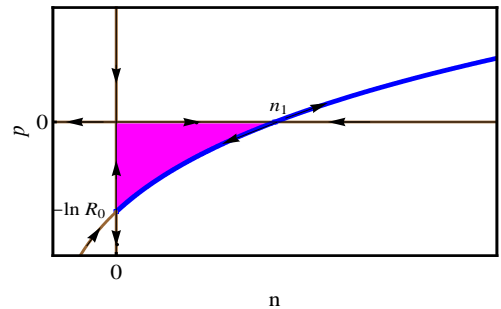}
\caption{(color online). Example of scenario A of population extinction driven by intrinsic noise. Shown are the zero-energy trajectories
of the
WKB Hamiltonian $H(n,p)$
for the reactions $A\stackrel{\lambda}{\rightarrow} 2A$,
$A\stackrel{\mu}{\rightarrow} \emptyset$ and
$2A\stackrel{\sigma}{\rightarrow} \emptyset$ \cite{kessler}.
The trajectories
denoted by the thicker line determine the WKB solution
for $\pi_n$, obtained in Ref. \cite{kessler}. The activation trajectory connects the metastable point $(n_1,0)$ and the fluctuational extinction point $(0,p_f)$, where $p_f=-\ln R_0$, and $R_0=\lambda/\mu$. The effective entropy barrier to extinction is equal to $N \Delta S$, where $N=\lambda/\sigma$ and $\Delta S$ is the area of the shaded region, given by Eq.~(\ref{Sexamp1}).}
\label{figB1}
\end{figure}
Here is an overview of the two extinction scenarios as described by the WKB approximation.
First, let $\bar{n}=0$ be a repelling fixed point of the rate equation, see Fig.~\ref{mf}a. In a stochastic description extinction occurs via a large fluctuation which, acting against an effective entropy barrier, brings the population
from a vicinity of $n=n_1$ directly to the
absorbing state $n=0$. In the WKB language this transition is possible because of the presence of the
fluctuational momentum $p=dS/dn$, see Fig.~\ref{figB1}.  The attracting and repelling fixed
points of the rate equation $\bar{n}=n_1$ and $\bar{n}=0$, respectively, become
hyperbolic fixed points of an extended phase plane $(n,p)$. Importantly, an additional hyperbolic fixed point $(0,p_f)$ - the fluctuational extinction point - appears here, with a zero coordinate, $n=0$, but a nonzero momentum $p_f$ \cite{Kamenev1,Kamenev2,AKM}.  The most probable path to extinction is the heteroclinic trajectory, directly
connecting the ``metastable point", that is the hyperbolic point $(n_1,0)$, and the ``fluctuational extinction point": the hyperbolic point $(0,p_f)$.  (Such escape trajectories - heteroclinic trajectories with a non-zero momentum  - are often called ``activation trajectories", see \textit{e.g.} \cite{dykman}.)
This is what we call extinction scenario A.

\begin{figure}
\includegraphics[width=2.6in,height=1.6in,clip=]{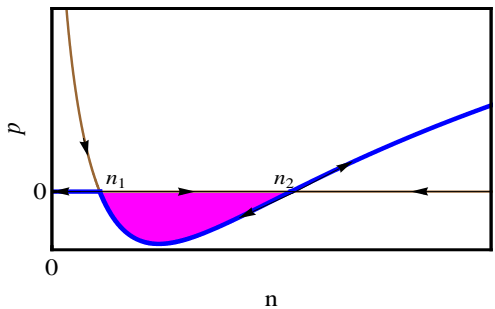}
\caption{(color online). Example of scenario B of population extinction driven by intrinsic noise. Shown are the zero-energy trajectories (\ref{zeroE}) on the phase plane $(n,p)$ of the WKB Hamiltonian (\ref{hamilex2}) for the reactions $A\stackrel{\mu}{\rightarrow} \emptyset$,
$2A\stackrel{\lambda}{\rightarrow} 3A$ and
$3A\stackrel{\sigma}{\rightarrow} 2A$.
The trajectories denoted by the thicker lines determine  the WKB solution for the
QSD. The most probable path to extinction first goes along the non-zero-momentum heteroclinic trajectory (the activation trajectory) which connects the points $(n_2,0)$ and $(n_1,0)$. Then the population flows almost deterministically to the extinction point $(0,0)$ along a zero-momentum segment (the relaxation trajectory). The effective entropy barrier to extinction is equal to $N \Delta S$, where $N=3\lambda/(2 \sigma)$, and  $\Delta S$ is the area of the shaded region, given by Eq.~(\ref{deltaS2}).}
\label{figB1a}
\end{figure}
Now let $\bar{n}=0$ be an attracting fixed point of the rate equation (\ref{rateeqgen}), so that the metastable population resides in the vicinity of $n=n_2$, see  Fig.~\ref{mf}b.  In the stochastic description, extinction occurs via a large fluctuation which brings the population
from a vicinity of $n=n_2$ to a vicinity of the \textit{repelling} fixed point $n_1$. From there the system flows into the
absorbing state $n=0$ ``downhill", that is almost deterministically. In the framework of WKB theory the transition from $n_2$
to $n_1$ occurs in the extended phase plane  $(n,p)$ where
all three fixed points are hyperbolic, see Fig.~\ref{figB1a}. Here the optimal path to extinction is composed of two segments: the non-zero-momentum
heteroclinic trajectory connecting the hyperbolic fixed points
$(n_2,0)$ and $(n_1,0)$ (the activation trajectory), and the zero-momentum segment going from $n=n_1$ to $n=0$ (the relaxation trajectory). This is what we call extinction scenario B.

The mean time to extinction (MTE) $\tau$ and/or the QSD $\pi_n$ of metastable single-species stochastic populations were calculated previously in particular examples in different contexts of physics, chemistry, population biology, epidemiology, cell biology, \textit{etc}. Among them there is a large body of work which approximated the master equation by an effective Fokker-Planck equation,
derived via the van Kampen system size expansion or related recipes. Once the  Fokker-Planck equation is obtained, the MTE and QSD can be calculated by standard methods \cite{kampen,gardiner}. Unfortunately, this approximation is in general uncontrolled. It fails in its description of
the tails of the QSD, and gives exponentially large errors in the MTE, as shown in Refs. \cite{gaveau,Doering,kessler,Assaf2}.

With a few exceptions, accurate analytic results for the MTE and QSD are only available for \textit{single-step} gain-loss processes: $r=\pm 1$ in Eq.~(\ref{master}). In this case the MTE can be determined exactly by employing the backward master
equation \cite{kampen,gardiner}. This yields a cumbersome analytic expression for the MTE which, for a large population size in the metastable state, can be simplified via a saddle-point approximation. Such a procedure was implemented in Ref.~\cite{Doering}. In its turn, the QSD $\pi_n$ of single-step processes can be calculated  from a recursive relation obtained when substituting Eq.~(\ref{qsdintro}) in the master equation. Several model examples of \textit{multi-step} processes were considered in Refs. \cite{turner,Assaf1,Assaf2,kessler}, all of them belonging to extinction scenario A.  We will generalize the previous results substantially and determine the MTE and QSD for quite a general set of gain-loss processes pertaining to extinction scenario A. We will also determine the MTE and QSD for extinction scenario B.

Our WKB theory starts with applying the ansatz (\ref{n30}) to an eigenvalue problem for the QSD $\pi_n$ which is nothing but the first excited eigenvector of the master equation. In the leading WKB order one arrives at the problem of finding zero-energy trajectories of an effective classical Hamiltonian \cite{dykman}.  There are two different types of zero-energy phase trajectories (in addition to the extinction line $q=0$): the activation and relaxation trajectories, which correspond to the fast and slow WKB modes, respectively \cite{MS}. To obtain the pre-exponents, one needs to consider the sub-leading WKB order. The WKB calculations are simpler for scenario A, as the relaxation trajectory does not play any role here. In scenario B both the activation, and the relaxation trajectories are important.  In both scenarios the WKB approximation  breaks down at $n={\cal O}(1)$. Here we find the QSD, up to a normalization constant, from a recursive relation, obtained by linearizing the process rates with respect to $n$ at sufficiently small $n$.  In scenario A it suffices to match the recursive solution with the fast-mode solution in their joint region of validity, in much the same way as it was done by Kessler and Shnerb \cite{kessler} in a particular example of three stochastic reactions. In Scenario B the slow mode dominates the WKB-solution at $n<n_1$. It diverges, however, at $n=n_1$. To obtain a regular solution there, one needs to go beyond the WKB approximation and account, in a close vicinity of $n=n_1$, for strong coupling between the fast- and slow-mode solutions. This can be done via the van-Kampen system size expansion of the master equation which does hold in the vicinity of $n=n_1$. This procedure was first implemented by Meerson and Sasorov \cite{MS}, in a model problem of noise-driven population explosion.  Then it has been employed by Escudero and Kamenev \cite{EsK} in the context of a WKB theory of stochastic population switches between two different metastable states. The theory of Escudero and Kamenev \cite{EsK} was formulated for quite a general set of gain-loss processes. In this paper we will adopt their general approach, and some of their notation, in the problem of population extinction.

Here is a plan of the remainder of the paper. Section~\ref{general} starts with a formulation of the eigenvalue problem for the QSD. Then we expose the WKB approximation and the fast- and slow-mode WKB solutions. The derivation here is quite general and holds for extinction scenarios A and B.
Section~\ref{recursion} presents a derivation of recursive solution of the quasi-stationary master equation for sufficiently small $n$. This derivation, which also holds for extinction scenarios A and B, is specific to population extinction. Except for simple particular cases (see \textit{e.g.} Ref. \cite{kessler}), it has not been attempted before. In Section \ref{sA} we match the fast-mode WKB solution with the recursive small-$n$ solution and obtain general expressions for the QSD and MTE in
scenario A.  In the same Section we obtain the QSD and MTE for single-step processes and near the transcritical bifurcation, characteristic of scenario A. Then we illustrate our theory on several particular examples, some of which investigated previously. In Section~\ref{sB} we determine the QSD and MTE for
scenario B. Then we again apply our results to single-step processes and near the saddle-node bifurcation, characteristic of scenario B. Furthermore, we consider a particular example of three stochastic reactions and compare our theoretical predictions with a numerical solution of the master equation. A summary of our results is presented in Section~\ref{conclusion}.

\section{Eigenvalue problem, WKB approximation, and fast- and slow-mode solutions}\label{general}

When starting at $t=0$ from a sufficiently large population, the probability distribution $P_n(t)$, as described by the master equation (\ref{master}) approaches, on a relaxation time scale $t_r$,  a long-lived metastable PDF peaked at a non-zero attracting fixed point of the rate equation. The metastable distribution is slowly ``leaking" to zero, see Eq.~(\ref{qsdintro}). Let us denote the non-zero attracting fixed point by $n=n_*$
($n_*=n_{1,2}$ for scenario A and B, respectively, see Figs. \ref{figB1} and \ref{figB1a}).  Using Eq.~(\ref{qsdintro}), we arrive at an eigenvalue problem for the QSD $\pi_n$, $n=1,2\dots$:
\begin{equation}\label{qsdmaster}
\sum_{r} \left[W_r(n- r)\pi_{n-r} -W_r(n)\pi_n\right] = -E \pi_n\,.
\end{equation}
Importantly,  the eigenvalue  $E =1/\tau$ turns out to be \textit{exponentially} small compared to the relaxation time $t_r$. Therefore, the term in the right hand
side of Eq.~(\ref{qsdmaster})
can be neglected \cite{dykman,kessler,MS,EsK}, and we have to deal with a quasi-stationary equation
\begin{equation}\label{qsdmaster1}
\sum_{r} \left[W_r(n- r)\pi_{n-r} -W_r(n)\pi_n\right] =0\,, \;n=1,2, \dots \,.
\end{equation}
For definiteness, we normalize the QSD to unity: $\sum_{n=1}^{\infty}\pi_n=1$. Once $\pi_n$ is found, we can use Eqs.~(\ref{master0}) and (\ref{qsdintro}) to calculate the MTE:
\begin{equation}\label{E}
E= 1/\tau = \sum_{r<0} W_r(-r) \pi_{-r}\,.
\end{equation}
Let us introduce a rescaled coordinate $q = n/N$, where $N\gg 1$ is the large parameter of the problem. The central assumption of our theory is  that, after a proper rescaling of time which will be introduced shortly,  the process rates can be represented as
\begin{equation}
\label{rateexp}
W_r(n) \equiv W_r(Nq) = Nw_r(q) + u_r(q) + {\cal O}(1/N)\,,
\end{equation}
where, for $q={\cal O}(1)$, $w_r(q)$ and $u_r(q)$ are ${\cal O}(1)$. This assumption guarantees that the population be long-lived, and is crucial both for the WKB-approximation that we present in this Section, and for the recursive solution of Eq.~(\ref{qsdmaster1}) that we will be dealing with later. As $q=0$ is the absorbing state, $w_r(0)=u_r(0)=0$.

For $n\gg 1$ we can employ the WKB ansatz (\ref{n30}):
\begin{equation}\label{fastmode}
\pi(q)\equiv \pi_{Nq} \simeq A e^{-NS(q) - S_1(q)}\,,
\end{equation}
where $S(q)$ and $S_1(q)$ are assumed to be ${\cal O}(1)$, and a constant prefactor $A$ is introduced for convenience, see below.
Now we assume that $|r_{max}| \ll N$, Taylor-expand  the functions of $q-r/N$ in Eq.~(\ref{qsdmaster1}) around $q$ and keep terms up to ${\cal O}(1)$ order.  We obtain the equation derived by Escudero and Kamenev \cite{EsK}:
\begin{eqnarray}\label{masterqsd}
&&\hspace{-9mm}\sum_r(Nw_r+u_r)\nonumber\\
&&\hspace{-9mm}\times\left[e^{rS^{\prime}}\left(1+\frac{r}{N}S_1^{\prime}-
\frac{r^2}{2N}S^{\prime\prime}-\frac{r}{N}\frac{w_r^{\prime}}{w_r}\right)-1\right]=0\,,
\end{eqnarray}
where the primes denote differentiation with respect to $q$. In the leading order ${\cal O}(N)$, this equation yields a
stationary Hamilton-Jacobi equation $H(q,S^{\prime})=0$, where
\begin{equation}\label{hamil}
H(q,p)=\sum_rw_r(q)\left(e^{rp}-1\right)
\end{equation}
is the effective Hamiltonian, and $p=S^{\prime}$ is the momentum \cite{dykman}. Therefore, in the leading WKB order, one needs to find \textit{zero-energy} phase trajectories of the Hamiltonian (\ref{hamil}). As $w_r(0)=0$ for any $r$, one such trajectory is $q=0$ at an arbitrary $p$: the extinction line. This line is of no importance in the WKB theory, however. What we need are phase trajectories $p = p(q)$.  One of them is the relaxation trajectory $p=0$. In general, there is one and only one additional phase trajectory for which $p=p_a(q) \neq 0$, except in some points $q$. Let us prove this statement. The Hamiltonian $H(q,p)$ vanishes at $p=0$. Differentiating Eq.~(\ref{hamil}) twice with respect to $p$, we obtain $H_{pp}(q,p)=\sum_r r^2 w_r(q) e^{rp}>0$. Therefore
$H$ is a convex function of $p$, and so it has one and only one additional real root \cite{nontrivial}. The relaxation trajectory $p=0$ and activation trajectory $p=p_a(q)$ give rise to the slow and fast WKB modes, respectively, as was shown in a particular example in Ref. \cite{MS}.

The $q$-dynamics along $p=0$ is described by the Hamilton's equation
\begin{equation}
\label{qdot}
\dot{q} = \frac{\partial H(q,p)}{\partial p}|_{p=0}\equiv H_p(q,0)= \sum_r r w_r(q)
\end{equation}
which is nothing but the (rescaled) rate equation~(\ref{rateeqgen}).
The nontrivial fixed points of the rate equation, $q_i=n_i/N$, are positive roots of the  equation $H_p(q_i,0)=0$. As $p_a(q_i)=0$,  the activation trajectory $p=p_a(q)$ crosses the relaxation trajectory in these fixed points. As $w_r(0)=0$ for any $r$, the small-$q$ expansion of Eq.~(\ref{qdot}) generically starts with a linear term in $q$. In the remainder of this paper we assume that the linear decay rate is non-zero in the leading order: $w_{-1}^{\prime}(0) =\alpha>0$. Therefore, we can always rescale time, and all the rates, by $\alpha$. This procedure uniquely defines the rescaling leading to Eq.~(\ref{rateexp}).

In extinction scenario A, see Fig.~\ref{figB1}, the most probable path to extinction is the heteroclinic trajectory
connecting the fixed point $(q_*,0)$  of the Hamiltonian system (here $q_*$ is the attracting point of the rate equation) with the fluctuational extinction point $(0,p_f)$.

In extinction scenario B, see Fig.~\ref{figB1a}, the most probable path to extinction is composed of two segments. The first one is the activation trajectory: a non-zero-momentum heteroclinic trajectory connecting the fixed point $(q_*,0)$ with an intermediate fixed point $(q_{rep},0)$, where $q_{rep}$ is a repelling fixed point of the rate equation. The second one is a relaxation segment $p=0$, connecting the point $(q_{rep},0)$ with the deterministic extinction point $(0,0)$.

For the fast mode one obtains \cite{dykman}
\begin{equation}\label{sfast}
S(q)=S^{(f)}(q)=\int^{q}d\xi\, p_a(\xi)\,,
\end{equation}
where the integration constant is already accounted for by the prefactor $A$ in Eq.~(\ref{fastmode}). For the slow mode $S(q)=S^{(s)}(q)=0$.

In the subleading ${\cal O}(1)$ order, Eq.~(\ref{masterqsd}) yields a first-order differential equation  for $S_1(q)$:
\begin{equation}
\sum_r w_r
e^{rp}\left(rS_1^{\prime}-\frac{r^2}{2}p^{\prime}-\frac{rw_r^{\prime}}{w_r}\right)+u_r(e^{rp}-1)=0\,,\label{s1calculation}
\end{equation}
where $p=p_a(q)$ for the fast mode, and $p=0$ for the slow mode.  It is convenient to use the identities
\begin{eqnarray}
&& H_q=\sum_r  w_r^{\prime}(q) (e^{rp}-1)\,,\;\;\; H_p=\sum_r r w_r(q) e^{rp}\,, \nonumber \\
&& H_{qq}= \sum_r  w_r^{\prime\prime}(q) (e^{rp}-1)\,,\;\;\;H_{pp}=\sum_r
r^2w_r(q)e^{rp}\,, \nonumber \\
&& H_{qp}=\sum_r rw_r^{\prime}(q)e^{rp}\,,
\label{identities}
\end{eqnarray}
where the subscripts $p$ and $q$ stand for the partial derivatives. To remind the reader, all the rates in Eqs.~(\ref{identities}) are rescaled
with respect to the linear decay rate constant $\alpha$. The fast-mode solution for $S_1(q)$ can be written as
\begin{eqnarray}
&& \hspace{-4mm} S_1^{(f)}(q)=\int^{q}d\xi\nonumber\\
&& \hspace{-4mm}\times\frac{H_{qp}(\xi,p_a)+\frac{1}{2}H_{pp}(\xi,p_a)p_a^{\prime}(\xi)
-\sum_ru_r(\xi)(e^{rp_a}-1)}{H_p(\xi,p_a)},\nonumber\\\label{s1fast}
\end{eqnarray}
where $p_a=p_a(\xi)$. This result was obtained by Escudero and Kamenev \cite{EsK} in the context of stochastic population switches. The quantity $H_p(\xi,p_a)$ in the denominator of the integrand in Eq.~(\ref{s1fast}) vanishes in every fixed point $\xi=q_i$ of the rate equation, including $\xi=0$.
To see how the integrand behaves at the fixed points, consider the equation $H[q,p_a(q)]=0$ for the activation
trajectory. Differentiating it with respect to $q$, we obtain
\begin{equation}\label{hp}
H_q[q,p_a(q)]+H_p[q,p_a(q)] p_a^{\prime}(q)=0\,.
\end{equation}
One more differentiation gives
\begin{equation}\label{hpq}
H_{qq}+H_p p_a^{\prime\prime}+(2H_{pq}+H_{pp}p_a^{\prime})p_a^{\prime}=0\,,
\end{equation}
evaluated at $p=p_a(q)$. By virtue of identities (\ref{identities}) each of the first two terms of Eq.~(\ref{hpq}) vanishes at $q=q_i$ and $p=p_a(q_i)=0$, so the expression
$2H_{pq}+H_{pp}p_a^{\prime}$ must also vanish there. As a result, the first two terms in the numerator of the integrand of Eq.~(\ref{s1fast}) cancel each other at $\xi=q_i$.  The remaining term in the numerator, $-\sum_ru_r(\xi)(e^{rp_a(\xi)}-1)$, is proportional to
$(\xi-q_i)$ in the vicinity of $\xi=q_i$, exactly as the quantity $H_p(\xi,p_a)$ in the denominator. Therefore, the integrand is
well behaved at $\xi=q_i>0$. At $q=0$, $S_1^{(f)}(q)$ diverges logarithmically, see below.
This divergence does not cause any concern, as the WKB approximation (which demands
$n\gg 1$, or $q\gg N^{-1}$, for its validity) does not hold for small $q$ anyway.

One can partially perform the integration over $q$ in Eq.~(\ref{s1fast}) by using Eqs.~(\ref{hp}) and (\ref{hpq}). After some algebra,
\begin{equation}
S_1^{(f)}(q)=- \ln \sqrt{|S^{\prime\prime}(q)|}+\Psi(q)\,,\label{S1final}
\end{equation}
where
\begin{equation}\label{Psi}
\Psi(q)=\int^{q}\left[\frac{H_{qq}(\xi,p_a)}{2H_q(\xi,p_a)}-\frac{\sum_ru_r(\xi)(e^{rp_a}-1)}{H_p(\xi,p_a)}\right]d\xi,
\end{equation}
and $S^{\prime\prime}(q)\equiv p_a^{\prime}(q)$. Note that $S^{\prime\prime}(q)$ does not vanish in any of the nontrivial fixed points $q_i$, so the logarithmic term in Eq.~(\ref{S1final}) is well behaved there \cite{q0}. Indeed, from Eq.~(\ref{hp})  $S^{\prime\prime}(q)= p_a^{\prime}(q)=-H_q[q,p_a(q)]/H_p[q,p_a(q)]$. By Taylor-expanding the numerator and  denominator in the vicinity of any nontrivial fixed point $q_i$, one can see that $S^{\prime\prime}(q_i)\neq 0$.

Therefore, the general fast-mode solution is given by Eq.~(\ref{fastmode}) with $S^{(f)}(q)$ from Eq.~(\ref{sfast}), and $S_1^{(f)}(q)$ from Eqs. (\ref{S1final}) and (\ref{Psi}). It also includes a constant prefactor $A=A_f$ which can be found immediately. Indeed, at $N \gg 1$ the QSD is strongly peaked around the attracting fixed point $q=q_*$. Here the fast-mode solution dominates, and $A_f$ can be found by normalizing to unity the gaussian asymptote of the QSD around $q=q_*$. The gaussian asymptote is obtained by expanding the QSD (\ref{fastmode}) in the vicinity of $q=q_*$:
\begin{equation}
\label{gauss}
\pi(q)\simeq A_f
e^{-NS^{(f)}(q_*)-S_1^{(f)}(q_*)-(N/2)S^{\prime\prime}(q_*)(q-q_*)^2}\,,
\end{equation}
where we have used the equalities $S^{\prime}(q_*)=p_a(q_*)=0$. To see that $S^{\prime\prime}(q_*)>0$,
one can again use Eq.~(\ref{hpq}). At $q=q_*$ it reads $2 H_{pq}(q_*,0)+H_{pp}(q_*,0) p_a^{\prime}(q_*)=0$.
By virtue of Eq.~(\ref{identities}) $H_{pp}(q_*,0)>0$. The quantity $H_{pq}(q_*,0)$ is
the $q$-derivative, evaluated at $q=q_*$, of the expression in the right hand side of the rate equation (\ref{qdot}). As the point $q=q_*$ is by assumption attracting, $H_{pq}(q_*,0)<0$. Therefore, $p_a^{\prime}(q_*)=S^{\prime\prime}(q_*)>0$, and the asymptote (\ref{gauss}) is indeed a gaussian distribution. Normalizing it to unity, we obtain
\begin{equation}
A_f=\sqrt{\frac{S^{\prime\prime}(q_*)}{2\pi N}}e^{NS^{(f)}(q_*)+S_1^{(f)}(q_*)},
\end{equation}
so the fast-mode solution is fully determined:
\begin{eqnarray}\label{fastmode1}
\pi(q) =\sqrt{\frac{S^{\prime\prime}(q_*)}{2\pi N}}e^{N[S^{(f)}(q_*)-S^{(f)}(q)]+S_1^{(f)}(q_*)-S_1^{(f)}(q)}\,,
\end{eqnarray}
with $S^{(f)}(q)$ from Eq.~(\ref{sfast}) and $S_1^{(f)}(q)$ from Eqs. (\ref{S1final}) and (\ref{Psi}).

Now consider the slow-mode solution for which $S^{(s)}(q)=0$. The subleading-order contribution $S_1^{(s)}(q)$ is found by putting
$p=0$ in Eq.~(\ref{s1fast}):
\begin{equation}
S_1^{(s)}(q) = \int^{q}d\xi\;\;\frac{H_{pq}(\xi,0)}{H_p(\xi,0)} = \ln H_p(q,0)\,.
\end{equation}
Then Eq.~(\ref{fastmode}) yields the general slow-mode solution for the QSD \cite{MS,EsK}:
\begin{equation}\label{slowmode}
\pi_s(q)=-\frac{A_s}{H_p(q,0)},
\end{equation}
where $A_s$ is an arbitrary constant. The minus sign is put here for convenience, because
in the region  of $0<q<q_1$, where the slow-mode solution is relevant (see Section~\ref{sB}), $\dot{q}=H_p(q,0)<0$ and $A_s>0$ . One can see from Eq.~(\ref{slowmode}) that the slow-mode solution
diverges in the fixed points of the rate equation \cite{MS}. This divergence will be cured in Section ~\ref{sB}.

We show in the following that, for a given extinction scenario and in a given region of $q$, only one of the modes,  either fast or slow,  dominates the resulting QSD, while the other one must be discarded.  Before we deal with this issue, however, we recall that the WKB approximation breaks down at $n={\cal O}(1)$. To find the QSD for \textit{all} $n$ we will solve Eq.~(\ref{qsdmaster1}) in the region of $n \ll N$ by recursion and then match the recursive solution with either the fast-mode (in scenario A), or the slow-mode (in scenario B) WKB solution in the joint region of their validity.

\section{Recursive solution}\label{recursion}
The objective of this Section is to approximately solve Eq.~(\ref{qsdmaster1}) at sufficiently small $n$. The exact criterion of smallness will appear later, when we match different solutions in joint regions of their validity.

In the leading order in $N$ we take $W_r(n)=N w_r(n/N)$, see Eq.~(\ref{rateexp}), and expand it in $n/N$ up to the linear term: $W_r(n)\simeq Nw_r(0)+n w_r^{\prime}(0) = n w_r^{\prime}(0)$. Then Eq.~(\ref{qsdmaster1}) becomes
\begin{equation}\label{QSDR1}
\sum_{r} w_r^{\prime}(0) \left[(n-r)\pi_{n-r}-n\pi_n\right]=0\,,
\end{equation}
where only processes with $w_r^{\prime}(0)\neq 0$ contribute.
One can look for particular solutions of this recursive equation in the form
$\pi_n=f_n/n$ thus arriving at an equation with $n$-independent coefficients:
\begin{equation}\label{QSDR2}
\sum_{r} w_r^{\prime}(0) (f_{n-r}-f_n)=0\,.
\end{equation}
In the remainder of this paper we make the following simplifying assumption:
\begin{equation}\label{wminus}
w_{-2}^{\prime}(0)=w_{-3}^{\prime}(0)= \dots =0.
\end{equation}
That is, we assume that the rates of the multi-step \textit{loss} processes $n\to n-m$, where $m=2,3,\dots$, do not have, in the leading order in $N$, linear terms in their Taylor expansion in $n$. This assumption is always satisfied for stochastic chemical reactions (where pairs, triplets, $\dots$, of reacting particles are needed to bring down the number of particles by $2,3, \dots$). The conditions~(\ref{wminus}) also hold for all models of population biology and epidemiology we are aware of.

Using  Eq.~(\ref{wminus}) and the equality $w_{-1}^{\prime}(0)=1$ (to remind the reader, we are using rescaled variables), we rewrite the recursive Eq.~(\ref{QSDR2}) as
\begin{equation}
f_{n+1}=\left[1+\sum_{r=1}^{K}w_r^{\prime}(0) \right] f_n- \sum_{r=1}^{K}w_r^{\prime}(0)
f_{n-r}\,, \label{QSDR3}
\end{equation}
where $K\equiv r_{max}\ll N$. If there is no degeneracy, the general solution of Eq.~(\ref{QSDR3}) is a linear combination of all particular solutions $f_n=\lambda^{-n}$, where $\lambda$ obeys the characteristic polynomial equation of degree $K+1$:
\begin{equation}\label{pol2}
\sum_{r=1}^K w_r^{\prime}(0) \lambda^{r+1}  - \left[1+\sum_{r=1}^K w_r^{\prime}(0)\right]
\lambda +1 =0\,.
\end{equation}
Note, that $\lambda=\lambda_0= 1$ is always a root. Let us show that Eq.~(\ref{pol2}) has one and only one additional positive root, $\lambda_1$, while all others roots $\lambda_2,\lambda_3,\dots,\lambda_K$ are either negative or complex. First, we establish a connection between the roots $\lambda_i$ of Eq.~(\ref{pol2}) and the crossing points with the $p$-axis of the zero-energy trajectories of the WKB Hamiltonian~(\ref{hamil}). By expanding $w_r(q\to 0)\simeq w_r^{\prime}(0)q$, Eq.~(\ref{hamil}) becomes
\begin{equation}\label{hamil2}
\sum_r w_r^{\prime}(0)\left(e^{rp}-1\right)=0,
\end{equation}
where only terms with $w_r^{\prime}(0)\neq 0$ contribute. Putting $e^{p}=\lambda$ and $w_{-1}^{\prime}(0)=1$ and using
Eq.~(\ref{wminus}), one can see  that Eq.~(\ref{hamil2}) coincides with Eq.~(\ref{pol2}). As we have shown that the equation $H(q,p)=0$ has, for any $q$, two and only two real solutions for $p$ \cite{nontrivial}, Eq.~(\ref{pol2}) also has two and only two real solutions for $\lambda$, both of them positive. The roots $\lambda_0=1$ and $\lambda_1$ correspond to the crossing points with the $p$-axis of the slow and fast modes, respectively.

Dividing
Eq.~(\ref{pol2}) by $1-\lambda$, we arrive at a polynomial equation of degree $K$:
\begin{equation}\label{pol3}
1-\sum_{r=1}^K w_r^{\prime}(0)(\lambda+\dots+\lambda^{r})=0\,.
\end{equation}
For $K\leq 4$ the roots of this polynomial can be expressed in radicals. For $K\geq 5$, they need to be computed numerically. Assume that we have found all of the roots of Eq.~(\ref{pol3}) $\lambda_i$, $i=1,2, \dots, K$. If there is no degeneracy,  the general solution for $f_n$ is
\begin{equation}\label{fn}
f_n=\sum_{i=0}^K C_i \lambda_i^{-n}=C_0+\sum_{i=1}^K C_i
\lambda_i^{-n}\,,
\end{equation}
where the coefficients $C_i$ are the following (see Appendix A for the derivation):
\begin{equation}\label{cires}
C_i=\frac{(-1)^K
f_1\displaystyle\prod_{j=1}^K\lambda_j}{\displaystyle\prod_{\begin{array}{c}j=0\\j\neq
i\\\end{array}}^K(\lambda_i-\lambda_j)}\,.
\end{equation}
The coefficient $C_0$, corresponding to the root $\lambda_0=1$, can be expressed through the coefficients $w_r^{\prime}(0)$, $r=1,2, \dots, K$ (see Appendix A):
\begin{equation}\label{c00main}
C_0=\frac{f_1}{1-w_1^{\prime}(0)-2w_2^{\prime}(0)-\dots-Kw_K^{\prime}(0)}\,.
\end{equation}

Before writing down the general solution of the recursive equation (\ref{QSDR1}) for the QSD, we recall a simple relation \cite{darroch} between $f_1$ and the MTE $\tau$. Using the rescaled reaction rates, we can rewrite Eq.~(\ref{E}) as
$$
\tau^{-1} = \alpha\sum_{r<0} W_r(-r)\,\pi_{-r}\,.
$$
In view of the conditions~(\ref{wminus}), only one term in the sum survives in the leading order.
As $w_{-1}^{\prime}(0)=1$, we obtain
$$
\tau^{-1} = \alpha W_{-1}(1) \pi_1 \simeq  \alpha \pi_1 = \alpha f_1\,,
$$
so $f_1 \simeq 1/(\alpha \tau)$. Now we switch to the rescaled variable $q=n/N$,
use the relation $\pi(q)=f_n/(Nq)$ and Eq.~(\ref{fn}), and obtain the small-$q$ asymptote of $\pi(q)$:
\begin{equation}\label{recsolution}
\pi(q)=\frac{(-1)^K \displaystyle\prod_{j=1}^K\lambda_j}{\alpha\tau Nq}\, \sum_{i=0}^K\frac{\lambda_i^{-Nq}}{\displaystyle\prod_{\begin{array}{c}j=0\\j\neq
i\\\end{array}}^K(\lambda_i-\lambda_j)}\,.
\end{equation}
The validity region of this asymptote (which includes the yet unknown $\tau$ to be found later) is scenario-dependent. It is a relatively narrow region $q\ll N^{-1/2}$ in scenario A, and a broader region $q\ll 1$ in scenario B. The difference comes from the fact that in scenario A the recursive solution needs to be matched, at $n \gg 1$, with a rapidly growing fast-mode solution, whereas in scenario B the matching needs to be done with a slowly varying slow-mode solution, see Sections \ref{sA} and \ref{sB}, respectively.

What is the role of complex roots of the polynomial equation (\ref{pol3}) in the recursive solution (\ref{recsolution})? These  can appear only for $K\geq 3$, and they come in complex conjugate pairs: $\lambda_j$ and  $\lambda_k=\overline{\lambda_j}$. One can show, by using Eq.~(\ref{cires}), that the coefficients $C_j$ and $C_k$, corresponding to $\lambda_j$ and $\overline{\lambda_j}$, are also complex conjugate: $C_k=\overline{C_j}$, so that $\pi(q)$ from Eq.~(\ref{recsolution}) is real-valued as expected. When complex roots are present, the QSD at small $n$ may exhibit rapidly decaying oscillations as a function of $n$.

Let us now determine the $n\gg 1$, or $q\gg N^{-1}$, asymptote of the QSD (\ref{recsolution}), in each of the two extinction scenarios. This asymptote will be matched, in each scenario, with the dominant WKB mode.
Equation~(\ref{recsolution}) includes $K+1$ terms. At $n\gg 1$ the leading contribution comes from the term with the smallest $|\lambda_i|$. The rest of the terms are exponentially small compared to the leading one and can be safely neglected.

In scenario A,  the two positive roots of Eq.~(\ref{pol2}) are $0<\lambda_1<1$ and $\lambda_0=1$, whereas the rest of the (negative or complex) roots obey the inequality
$|\lambda_{i>1}|>\lambda_1$, see Appendix B. In this case the asymptote of the recursive solution (\ref{recsolution}) at $n\gg 1$, or $q \gg N^{-1}$, is
\begin{equation}\label{rec1}
\pi(q)\simeq \frac{A_1\lambda_1^{-Nq}}{\alpha\tau Nq},
\end{equation}
where the positive constant $A_1$ (see Appendix B) satisfies
\begin{equation}\label{c1}
A_1=\frac{(-1)^K \displaystyle\prod_{j=1}^K\lambda_j}{\displaystyle\prod_{\begin{array}{c}j=0\\j\neq
1\\\end{array}}^K(\lambda_1-\lambda_j)}\,.
\end{equation}
In this case  $\lambda_1=e^{p_f}$ corresponds, in the WKB-language, to the $p_f<0$ crossing point of the activation trajectory and the $p$-axis, see Fig.~\ref{figB1}. Therefore, to set the ground for matching Eq.~(\ref{rec1}) with the WKB solution in scenario A, we can rewrite Eq.~(\ref{rec1}) as
\begin{equation}\label{rec11}
\pi(q)\simeq \frac{A_1}{\alpha\tau N}\frac{e^{-N p_{f} q}}{q}.
\end{equation}

In scenario B the $n \gg 1$ asymptote of Eq.~(\ref{recsolution}) is quite different. Here the root of Eq.~(\ref{pol2}) with the smallest absolute value is $\lambda_0=1$, see Appendix C. Therefore, the $i=0$ term in Eq.~(\ref{recsolution}) is dominant, and we obtain
\begin{equation}\label{rec2}
\pi(q)\simeq \frac{1}{\alpha\tau Nq\left[1-w_1^{\prime}(0)-2w_2^{\prime}(0)-\dots-K w_K^{\prime}(0)\right]}\,,
\end{equation}
where we have used Eqs.~(\ref{cires}) and (\ref{c00main}). The asymptote~(\ref{rec2}) can be expressed in terms of the WKB Hamiltonian~(\ref{hamil}). Using Eq.~(\ref{identities}), we obtain
$H_{qp}(0,0)=\sum_r rw_r^{\prime}(0)$. Recalling that $w_{-1}^{\prime}(0)=1$ and using Eq.~(\ref{wminus}),
we can rewrite $H_{qp}(0,0)$  as
\begin{equation}\label{hpqR}
H_{qp}(0,0)=Kw_{K}^{\prime}(0)+(K-1)w_{K-1}^{\prime}(0)+\dots+w_{1}^{\prime}(0)-1\,.
\end{equation}
As $q=0$ is an attracting point here, $H_{qp}(0,0)<0$.
Then, using Eq.~(\ref{hpqR}), the asymptote~(\ref{rec2}) becomes
\begin{equation}\label{rec22}
\pi(q)\simeq \frac{1}{\alpha\tau N|H_{qp}(0,0)| q}\,.
\end{equation}
Note that  $\lambda_0=1=e^{p_s}$ corresponds  to the zero-momentum ($p_s=0$) crossing point of the relaxation trajectory and the $p$-axis, see Fig.~\ref{figB1a}.

\section{Extinction scenario A}\label{sA}
\subsection{General case: multi-step processes}
In this section we calculate the MTE and QSD for extinction scenario A.
Here extinction occurs along the activation trajectory: the heteroclinic trajectory,
connecting the metastable point $(n_1,0)$ and the fluctuational extinction point $(0,p_f)$ of the phase plane $(n,p)$, see Fig.~\ref{figB1}. In this case the slow-mode solution is negligible compared to the fast-mode solution in the entire region of $q>0$. Furthermore, the fast-mode solution (\ref{fastmode1}) can be directly matched with the recursive solution (\ref{recsolution}) in the joint region of their validity which turns out to be $1\ll n\ll N^{1/2}$, or $N^{-1}\ll q\ll N^{-1/2}$.

To implement the matching procedure, we first find the $q\ll N^{-1/2}$ asymptote of the
fast-mode solution (\ref{fastmode1}). Because of the divergence of $S_1^{(f)}(q)$ at $q=0$, we should proceed with care. Let us rewrite Eq.~(\ref{fastmode1}) as
\begin{equation}\label{fm1}
\pi(q)= \frac{\sqrt{S^{\prime\prime}(q_1)}}{\sqrt{2\pi N} q}\, e^{N[S^{(f)}(q_1)-S^{(f)}(q)]+S_1^{(f)}(q_1)-[S_1^{(f)}(q)-\ln q]}\,.
\end{equation}
Here we have introduced the $1/q$ prefactor which diverges at $q=0$, and made up for it by adding $\ln q$ in the exponent. Let us show that the expression $S_1^{(f)}(q)-\ln q$ in the exponent is regular at $q=0$.
We represent $\ln q$ as $\int^q d\xi/\xi$ and use Eq.~(\ref{s1fast}) to rewrite  $S_1^{(f)}(q)-\ln q$ as an integral over $\xi$.
Now we Taylor-expand the integrand
$$
\frac{H_{pq}(\xi,p_a)+\frac{1}{2}H_{pp}(\xi,p_a)p_a^{\prime}(\xi)
-\sum_ru_r(\xi)(e^{rp_a}-1)}{H_p(\xi,p_a)}-\frac{1}{\xi}
$$
in the vicinity of $\xi=0$ up to linear terms. The divergent terms cancel out, and the remaining expression
\begin{equation}
\frac{(1/2)H_{ppq}(0,p_f)p_a^{\prime}(0)-\sum_ru_r^{\prime}(0)(e^{rp_f}-1)}{H_{pq}(0,p_f)}\,,\label{regular}
\end{equation}
is finite. Now we rewrite Eq.~(\ref{fm1}) as
\begin{equation}\label{fm3}
\pi(q)=\frac{\sqrt{S^{\prime\prime}(q_1)}}{\sqrt{2\pi N}}\frac{q_1}{q}e^{N[S^{(f)}(q_1)-S^{(f)}(q)]+\phi(q_1)-\phi(q)}\,,
\end{equation}
where
\begin{equation}\label{phi}
\phi(q)= S_1^{(f)}(q)-\ln q
\end{equation}
is regular at $q=0$.
By Taylor-expanding the exponent of Eq.~(\ref{fm3}) around $q=0$ to first order, we obtain  the $q\ll N^{-1/2}$ asymptote of the
fast-mode solution:
\begin{eqnarray}\label{fm4}
\pi(q)\simeq \frac{\sqrt{S^{\prime\prime}(q_1)}}{\sqrt{2\pi N}}\frac{q_1}{q}e^{-Nqp_f}\,e^{N[S^{(f)}(q_1)-S^{(f)}(0)]+\phi(q_1)-\phi(0)}.\nonumber\\
\end{eqnarray}
This asymptote can be matched with the asymptote of the recursive solution at $N^{-1}\ll q \ll N^{-1/2}$, given by Eq.~(\ref{rec11}). This matching yields
\begin{eqnarray}
\tau=\frac{A_1\sqrt{2\pi}}{\alpha q_1\sqrt{N S^{\prime\prime}(q_1)}}e^{N[S^{(f)}(0)-S^{(f)}(q_1)]+\phi(0)-\phi(q_1)},\label{tau1}
\end{eqnarray}
where $\phi(q)$ is given by Eq.~(\ref{phi}),  $\alpha$ is the linear decay rate constant in physical units, and $A_1$ is given by Eq.~(\ref{c1}) \cite{physical}. The general expression (\ref{tau1}) for the MTE in scenario A is one of the main results of this work. The leading term in the
exponent, proportional to $N$, is the effective entropy barrier to extinction. The proportionality factor is the absolute value of the area under the activation trajectory, see an example in Fig.~\ref{figB1} \cite{Kamenev1}. Noticeable is the presence of the large factor $N^{1/2}$ in the pre-exponent.
The constant $A_1$ has a clearly non-WKB nature, as it comes from the recursive solution of the quasi-stationary master equation at small $n$ and is contributed to by \textit{all} of the roots $\lambda_i$, $i=0,1, \dots, K$.

Another important result is the QSD in extinction scenario A. It is determined by the asymptotes~(\ref{fastmode1}) and (\ref{recsolution}) which coincide, in the leading order,
in their joint region of validity  $N^{-1}\ll q\ll N^{-1/2}$, or  $1\ll n\ll N^{1/2}$.

\subsection{Single-step processes}
Remaining within scenario A, we now turn to an important sub-class of stochastic population processes: single-step processes. Here there are only two non-zero process
rates: $W_{\pm 1}(Nq)\equiv W_{\pm}(Nq)= Nw_{\pm}(q)+u_{\pm}(q)+\dots$,
where all the rates are normalized by the linear decay rate constant $w_{-}^{\prime}(0)$.
In this case the expressions for the MTE and QSD can be simplified considerably. The WKB Hamiltonian (\ref{hamil}) becomes
\begin{equation}\label{singleHam}
H(q,p)=w_+(q)(e^p-1)+w_{-}(q)(e^{-p}-1)\,.
\end{equation}
The  rate equation is $\dot{q}=w_+(q)-w_-(q)$. In scenario A one has $w_+^{\prime}(0)>w_-^{\prime}(0)= 1$. Here it is convenient to denote the ratio of the linear birth and death rates by $R\equiv w_+^{\prime}(0)/w_-^{\prime}(0)=w_+^{\prime}(0)$. For $R>1$ the fixed point $q=0$ of the rate equation is repelling.  The activation trajectory is
$p_a(q)=-\ln [w_{+}(q)/w_-(q)]$, and
\begin{equation}\label{Ssingle}
S^{(f)}(q)=-\int^q\ln \frac{w_{+}(\xi)}{w_-(\xi)}\,d \xi.
\end{equation}
Now we calculate the following quantities on the activation trajectory:
\begin{eqnarray}
&&p_a^{\prime}(q)=S^{\prime\prime}(q)=\frac{w_{-}^{\prime}}{w_{-}}-\frac{w_+^{\prime}}{w_+}\,,\;\;H_p(q,p_a)=w_{-}-w_+\,,\nonumber\\
&&H_{pq}(q,p_a)\!=\!\frac{w_{-}w_+^{\prime}}{w_{+}}\!-\!\frac{w_{+}w_{-}^{\prime}}{w_{-}},\;
H_{pp}(q,p_a)\!=\!w_{-}\!+\!w_+,\nonumber\\
&&\sum_r u_r (e^{rp_a}\!-\!1)\!=\!u_+\left(\frac{w_{-}}{w_+}-1\right)\!+\!u_{-}\left(\frac{w_{+}}{w_{-}}-1\right).\label{identities1}
\end{eqnarray}
Substituting these into Eq.~(\ref{s1fast}), we
obtain after simplifications
\begin{eqnarray}\label{S11}
\hspace{-2mm}S_1^{(f)}(q)=\int^q \left(\frac{u_-}{w_-}-\frac{u_{+}}{w_{+}}\right)d\xi+\frac{1}{2}\ln\left[w_+(q) w_-(q)\right]\!.
\end{eqnarray}
Plugging this into Eq.~(\ref{phi}) yields
\begin{equation}\label{phi1}
\phi(q)=\int^q \left(\frac{u_-}{w_-}-\frac{u_{+}}{w_{+}}\right)d\xi+\frac{1}{2}\ln\left[\frac{w_+(q) w_-(q)}{q^2}\right]\,.
\end{equation}
Now we can calculate $e^{\phi(0)-\phi(q_1)}$ which enters Eqs.~(\ref{fm4}) and (\ref{tau1}):
\begin{equation}\label{phi2}
e^{\phi(0)-\phi(q_1)}=\frac{q_1\sqrt{R}}{w_+(q_1)}\exp\left[\int_{0}^{q_1} \left(\frac{u_{+}}{w_{+}}-\frac{u_-}{w_-}\right)dq\right],
\end{equation}
where we have used the following relations: (i) as $q\to 0$, $w_+(q) w_-(q)/q^2\to  w_+^{\prime}(0)w_-^{\prime}(0)=R$, (ii)  $w_+(q_1)=w_-(q_1)$ (as $q_1$ is a fixed point of the rate equation), and (iii) $w_{-}^{\prime}(0)=1$ because of the rescaling of the rates.

Now we turn to the recursive solution at small $n$, presented in Section \ref{recursion}.  For single-step processes $K=1$, and so Eq.~(\ref{pol2}) has only two roots: $\lambda_0=1$ and $\lambda_1=1/w_{+}^{\prime}(0)=1/R$.
Therefore, rewriting the small-$q$ asymptote (\ref{recsolution}) of the QSD in terms of $\pi_n$, we obtain
\begin{equation}%\label{}
\pi_n=\frac{(R^n-1) f_1}{(R-1)n}\,,
\end{equation}
whereas the constant $A_1$ from Eq.~(\ref{c1}) is $1/(R-1)$. Plugging this constant and Eqs.~(\ref{Ssingle}) and
(\ref{phi2}) in Eq.~(\ref{tau1}) for the MTE, we obtain
\begin{equation}
\tau=\frac{\sqrt{2\pi\,R}\,\,e^{\int_{0}^{q_1}\left(\frac{u_{+}}{w_{+}}-\frac{u_-}{w_-}\right)dq}}
{\alpha(R-1)\, w_+(q_1)\sqrt{N S^{\prime\prime}(q_1)} }\,e^{N\int_{0}^{q_1}\ln \left( \frac{w_{+}}{w_-}\right)dq}\,.\label{tau11}
\end{equation}
In the particular case $u_{+}=u_{-}=0$ Eq.~(\ref{tau11}) coincides with Eq.~(19) of Ref.~\cite{Doering}, obtained, via a saddle-point approximation, from the exact expression for the MTE of a single-step process. Doering \textit{et al.} assumed in their derivation that the subleading contributions $u_{\pm}$ to the process rates $W_{\pm 1}(n)$, see Eq.~(\ref{rateexp}),  vanish. As a result, the factor $\exp\left[\int_{0}^{q_1}\left(u_{+}/w_{+}-u_-/w_-\right)dq\right]$
is absent from their Eq.~(19). While the assumption  $u_{+}=u_{-}=0$ may hold in some simple models, it does not hold in general. For example, it does not hold for stochastic chemical reactions where the rates are combinatorial, as in one of the examples we present in subsection \ref{sA}D below.

\subsection{Extinction near transcritical bifurcation point}
Now let us return to a general set of (not necessarily single-step) processes. Our objective is to simplify the MTE (\ref{tau1}) in the special regime when the population, as described by the rate equation, is very close to the characteristic (transcritical) bifurcation point of scenario A. Here the attracting point $q=q_1$ is very close to the repelling point $q=0$, so that $q_1\ll 1$. This  also implies $|p_f|\ll 1$ \cite{Kamenev2,AKM1}.  Taylor-expanding Eq.~(\ref{hamil}) in $q$ and $p$ around $q=p=0$, we obtain
\begin{equation}
H(q,p)\simeq qp\sum_r \left[r w_r^{\prime}(0)+\frac{q}{2}r w_r^{\prime\prime}(0)+\frac{p}{2}r^2 w_r^{\prime}(0)\right]=0.
\end{equation}
The trivial solutions are the extinction line $q=0$ and the relaxation trajectory $p=0$, whereas the nontrivial solution yields a straight-line activation trajectory.  Using Eq.~(\ref{identities}) and expanding the algebraic equations $H_p(q,0)=0$ for $q_1$, and $H_q(0,p)=0$ for $p_f$ at small $q$ and $p$, we can represent the activation trajectory as
\begin{equation}\label{actbif1}
p_a(q)=-p_f \left(\frac{q}{q_1}-1\right)\,.
\end{equation}
Here
\begin{equation}\label{q1bif}
q_1=-\frac{2H_{qp}(0,0)}{H_{qqp}(0,0)}\,,
\end{equation}
where $H_{qqp}(0,0)<0$, and
\begin{equation}\label{pfbif}
p_f=-\frac{2H_{qp}(0,0)}{H_{qpp}(0,0)}\,,
\end{equation}
where $H_{qpp}(0,0)>0$. Exactly at the bifurcation the rate constants are such that $H_{qp}(0,0)=0$. Here
the attracting fixed point $q_1$ merges with the repelling point $q=0$.  The coordinate of the attracting fixed point  $q_1\equiv\delta$ can serve here as the distance to the bifurcation. [The third derivatives of the Hamiltonian, which appear in the denominators of Eqs.~(\ref{q1bif}) and (\ref{pfbif}), are generically of order unity.]

\begin{figure}
\includegraphics[width=2.6in,height=1.6in,clip=]{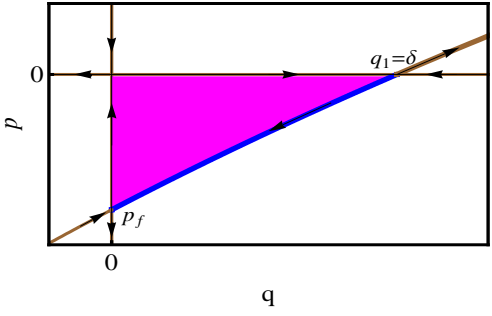}
\caption{(color online). Shown are typical zero-energy trajectories of the
WKB Hamiltonian (\ref{hamil}) in scenario A, close to the bifurcation where $q_1=\delta\ll 1$, and $p_f\ll 1$.
Here, the activation trajectory is a straight line, and $\Delta S$ given by Eq.~(\ref{triangle}) is the area of the shaded triangle.} \label{bifA}
\end{figure}

Now, using Eq.~(\ref{actbif1}), we can calculate $S^{\prime\prime}(q_1)=p_a^{\prime}(q_1)=-p_f/q_1=-p_f/\delta>0$,
the fast-mode action
\begin{equation}\label{actionbif}
S^{(f)}(q)=\int^q p_a(\xi)d\xi=p_f q-\frac{p_f}{\delta}\frac{q^2}{2}\,,
\end{equation}
and the accumulated action between the points $q=q_1=\delta$ and $q=0$
\begin{equation}\label{triangle}
\Delta S=S^{(f)}(0)-S^{(f)}(q_1)=-\frac{p_f q_1}{2}=-\frac{p_f \delta}{2}>0\,.
\end{equation}
This quantity is the area of a triangle \cite{Kamenev2}, see Fig.~\ref{bifA}.

To find the fast-mode correction to the action, $S_1^{(f)}(q)$, we expand Eq.~(\ref{s1calculation}) in the vicinity of $q=0$ and $p=0$, keeping only the leading order terms. This yields
\begin{equation}
\left[qS_1^{\prime}(q)-1\right] \sum_r r w_r^{\prime}(0)\simeq 0\,,
\end{equation}
whereas the subleading terms $u_r$ in the rate expansion do not contribute. The solution of this equation is $S_1^{(f)}(q)=\ln q$. Then, by virtue of Eq.~(\ref{phi}), we obtain $\phi(q)=0$. That is, near the bifuraction, the subleading WKB correction vanishes.

Now let us consider the coefficient $A_1$ [see Eq.~(\ref{c1})] which enters Eq.~(\ref{tau1}).
Among the roots $\lambda_i$ of the polynomial (\ref{pol2}), which contribute to $A_1$, two are special: $\lambda_0=1$ and
$\lambda_1=e^{p_f}$. Near the bifurcation $|p_f| \ll 1$, so we can write $\lambda_1 \simeq 1+p_f<1$. As a result, $A_1=\tilde{A}_1/p_f$. where
\begin{equation}\label{A1bif}
\tilde{A}_1=\frac{(-1)^K \displaystyle\prod_{j=2}^K\lambda_j}{\displaystyle\prod_{j=2}^K(1-\lambda_j)}
\end{equation}
is a negative constant of order unity, and we have put $\lambda_1= 1$ in the expression for $\tilde{A}_1$. Furthermore, the roots $\lambda_2, \lambda_3, \dots, \lambda_K$ of the polynomial (\ref{pol3}) can be evaluated \textit{at} the bifurcation point.  As a result, one can express $\tilde{A}_1$ via the linear branching rates $w_r^{\prime}(0)$. After some algebra
\begin{equation}\label{A1bif1}
\tilde{A}_1=-2\left[\displaystyle\sum_{r=1}^K r(r+1)w_r^{\prime}(0)\right]^{-1}\,.
\end{equation}
Substituting all of the above into Eq.~(\ref{tau1}),
we obtain the MTE close to the bifurcation point:
\begin{equation}\label{taubif}
\tau=\sqrt{\frac{2\pi}{N}}\frac{|\tilde{A}_1|}{\alpha D_A^3\,\delta^2}
\exp\left(\frac{N D_A^2\delta^2}{2}\right)\,,
\end{equation}
where $\tilde{A}_1$ and $D_A = \sqrt{|H_{qqp}(0,0)|/H_{qpp}(0,0)}={\cal O}(1)$ should be evaluated \textit{at} the bifurcation.  Equation~(\ref{taubif}) is valid when $N\delta^2\gg 1$. For sufficiently large $N$ this strong inequality is compatible with the strong inequality  $\delta \ll 1$ which describes closeness to the bifurcation. Note that the constant $\tilde{A}_1={\cal O}(1)$  is determined by the full small-$n$ recursive solution that we found in section \ref{recursion}.

Although Eq.~(\ref{tauuni}) breaks down at $\delta \sim  N^{-1/2}$, one can still
predict a scaling relation for the MTE in this region: $\tau \sim N^{1/2}/\alpha$.
The symbol $\sim$, here and in the following, means ``of the same order as".

\subsection{Examples}
We will now illustrate our theory by calculating the MTE in four pedagogical examples of extinction scenario A. The first three of them are single-step processes: the logistic Verhulst model of population dynamics, a set of three chemical reactions, and the SIS model of epidemics. The fourth example - another set of three chemical reactions - involves a two-step process. We will also consider all of these examples near the bifurcation.

\subsubsection{Verhulst model}
The generalized Verhulst model is a stochastic logistic model: a single-step Markov process with birth and death rates
\begin{eqnarray}
W_{+1}&\equiv& W_+=\alpha_1 n-\alpha_2 n^2\nonumber\\
W_{-1}&\equiv& W_-=\beta_1 n+\beta_2 n^2,\label{logrates}
\end{eqnarray}
respectively, where $\alpha_1,\alpha_2,\beta_1$ and $\beta_2$ are non-negative rate constants. The quadratic corrections account for competition for resources  \cite{Nasell}. It is customary to put $\alpha_2=0$ in Eq.~(\ref{logrates}) \cite{Doering}, and this is what we will do here. Rescaling time by the linear death rate constant $\beta_1$, we bring the rates to the form given by Eq.~(\ref{rateexp}): $W_+=B n$ and $W_-=n+Bn^2/N$, where $B=\alpha_1/\beta_1$ is the ratio of the linear birth and death rates, and $N=B\beta_1/\beta_2$. According to Eq.~(\ref{rateexp}) $w_+=B q$, $w_-=q+Bq^2$, and $u_+=u_-=0$.
At $B>1$ the fixed point $q=0$ of the rate equation is repelling, whereas
$q_1=1-1/B>0$ is attracting. Here we have $S^{\prime\prime}(q_1)=p_a^{\prime}(q_1)=1$, $A_1=1/(B-1)$ and
$$\int_0^{q_1}\ln \frac{w_-}{w_+}dq=\frac{1-B+\ln B}{B}.$$
Therefore, the MTE [Eq.~(\ref{tau11})] in physical time units is
\begin{equation}\label{tauver}
\tau=\frac{1}{\beta_1}\sqrt{\frac{2\pi}{N}}\frac{\sqrt{B}}{(B-1)^2}\exp\left[N\left(\frac{B-1- \ln B }{B}\right)\right]\,,
\end{equation}
which coincides with previous results obtained by different methods \cite{Doering,Nasell}.

In this simple example the process rates satisfied the conditions $u_{+}=u_{-}=0$, so Eq.~(\ref{tauver}) could have been obtained from Eq.~(19) of Ref.~\cite{Doering}. In the next example we relax one of the two conditions and show that, as predicted by our more general Eq.~(\ref{tau11}), the pre-exponent of the MTE changes.

\subsubsection{Reactions $A\rightleftarrows2A$ and $A\to 0$}
Consider a set of three reactions among particles $A$: branching $A\stackrel{\lambda}{\rightarrow} 2A$,  a reverse reaction $2A\stackrel{\sigma}{\rightarrow} A$, and decay $A\stackrel{\mu}{\rightarrow} \emptyset$. As observed in Ref. \cite{AKM1}, this set of reactions can be viewed as a generalization of the Verhulst model considered in the previous example. Indeed, by imposing a special relation, $\sigma=2(\mu-1)$, between the rate constants, and by denoting $\mu=1+B/N$ and $\lambda=B$, one recovers the process rates $w_{\pm}$ and $u_{\pm}$ of the Verhulst model. For this special choice of rate constants one has $u_{+}=u_{-}=0$ which yields Eq.~(\ref{tauver}) for the MTE. Now, what if the rate constants $\sigma$ and $\mu$ are independent? As usual, we can normalize time and the reaction rates by $\mu$ and denote $\tilde{B}= \lambda/\mu$ and $\tilde{N}= 2\lambda/\sigma$. By virtue of Eq.~(\ref{rateexp}) we can write $w_+=\tilde{B} q$, $w_-=q+\tilde{B}q^2$, $u_+=0$ and $u_-=\tilde{B}q$.   The rescaled rate constants are identical to those of the Verhulst model, except that $u_-$ is now nonzero. As a result,
$$\exp\left[\int_{0}^{q_1}\left(\frac{u_{+}}{w_{+}}-\frac{u_-}{w_-}\right)dq\right]=\tilde{B}\,,$$
and Eq.~(\ref{tau11}) for the MTE yields
\begin{equation}\label{tauvergen}
\tau=\sqrt{\frac{\pi\sigma}{\mu}}\frac{\lambda}{(\lambda-\mu)^2}
\exp\left[\frac{2}{\sigma}\left(\lambda-\mu+\mu\ln\frac{\mu}{\lambda}\right)\right]\,,
\end{equation}
where we have returned to physical units.  This result cannot be obtained from Eq.~(19) of Ref.~\cite{Doering}.

\subsubsection{SIS model}
Now let us consider the well-known SIS model of epidemics, see \textit{e.g.} Refs. \cite{Nasell,Ovas} and references therein. The SIS model deals with dynamics of a population which consists of two groups of individuals: susceptible to infection and infected. It is assumed that infection does not confer any long-lasting immunity, and infected individuals become susceptible again after infection. When demography (births and deaths) is negligible,  the total number $N$ of individuals in the two groups is conserved. As a result, the model becomes effectively single-population, with the effective rates
$$
W_+=\lambda n(N-n)\;,\;\;\;W_-=\mu n\,.
$$
Mathematically, this model is just another example of the generalized Verhulst model, see Eq.~(\ref{logrates}), where one chooses $\beta_1 \neq 0$ but $\beta_2=0$ \cite{Nasell}.

Let us denote $R_0=\lambda N/\mu$ and rescale time and rates by the linear decay rate constant $\mu=\lambda N/R_0$. The rescaled rates become $w_+=R_0(q-q^2)$ and $w_-=q$, while $u_+=u_-=0$.
The fixed point $q_1=1-1/R_0$ of the rate equation is attracting when $R_0>1$. Furthermore, $S^{\prime\prime}(q_1)=p_a^{\prime}(q_1)=R_0$, and $A_1=1/(R_0-1)$ (as in the above notation $R=R_0$). Finally, $$\int_0^{q_1}\ln \frac{w_-}{w_+}dq=1 - \frac{1}{R_0} - \ln R_0.$$
Therefore, the MTE [Eq.~(\ref{tau11})], in physical time units, is given by
\begin{eqnarray}\label{tausis}
\tau=\frac{1}{\mu}\sqrt{\frac{2\pi}{N}}\frac{R_0}{(R_0-1)^2}\exp\left[N\left(\ln R_0+\frac{1}{R_0}-1\right)\right]\,,
\end{eqnarray}
which coincides with previous results obtained by different methods \cite{Nasell,Ovas,diffR0}.

\subsubsection{Branching-annihilation-decay}
Now we consider another set of stochastic reactions among particles $A$ which include, in addition to single-step processes  $A\stackrel{\lambda}{\rightarrow} 2A$ and $A\stackrel{\mu}{\rightarrow} \emptyset$,  a two-step process: binary annihilation
$2A\stackrel{\sigma}{\rightarrow} \emptyset$. This problem was previously solved by
Kessler and Shnerb \cite{kessler}. Here we show that their result for the MTE follows from our Eq.~(\ref{tau1}).

In our notation, the transition rates between the states $n$ and $n+r$ are given by
\begin{equation}
W_1=\lambda n\;,\;W_{-1}=\mu n\,,\;\mbox{and}\;W_{-2}=\frac{\sigma n(n-1)}{2}\,.
\end{equation}
Rescaling time $\mu t \to t$ and denoting $R_0=\lambda/\mu$ and $N=\lambda/\sigma$, we obtain Eq.~(\ref{rateexp}) with \begin{eqnarray}
w_1&=&R_0 q\;,\;\;\;w_{-1}=q\;,\;\;\;w_{-2}=\frac{R_0 q^2}{2},\nonumber\\
u_{1}&=&0\;,\;\;\;u_{-1}=0\;,\;\;\;u_{-2}=-\frac{R_0 q}{2}.
\end{eqnarray}
In the rescaled notation, the attracting fixed point is $q_1=1-1/R_0$ which demands $R_0>1$.
The WKB Hamiltonian (\ref{hamil}) takes the form
\begin{equation}\label{hamilex1}
H(q,p)=R_0 q\left(e^p-1\right)+q\left(e^{-p}-1\right)+\frac{R_0 q^2}{2}\left(e^{-2p}-1\right)\,.
\end{equation}
Solving the equation $H[q,p_a(q)]=0$, we obtain the activation trajectory
\begin{equation}
p_a(q)=S^{\prime}(q)=\ln\left(\frac{u+v}{4R_0}\right)\,,
\end{equation}
where $u=2+qR_0$ and $v=\sqrt{u^2+8qR_0^2}$. The zero-energy phase trajectories of this system are
shown in Fig.~\ref{figB1}.

Now we use Eqs.~(\ref{sfast}), (\ref{s1fast}), and (\ref{phi}) and obtain
\begin{eqnarray}\label{Sexamp1}
&&\Delta S=S^{(f)}(0)-S^{(f)}(q_1)\nonumber\\
&&=2\left\{1-\frac{1}{R_0}
+\left(1+\frac{1}{R_0}\right)\ln \left[\frac{1}{2}\left(1+\frac{1}{R_0}\right)\right]\right\}
\end{eqnarray}
and
\begin{equation}
e^{\phi(0)-\phi(q_1)}=\frac{2R_0}{\sqrt{(R_0+1)(3R_0-1)}}\,.
\end{equation}
Furthermore, $S^{\prime\prime}(q_1)=2R_0/(3R_0-1)$ and, as in the previous examples, the only root of Eq.~(\ref{pol3}) is $\lambda_1=1/R_0$. Therefore,
by using Eq.~(\ref{c1}), we have $A_1=1/(R_0-1)$. Substituting all of the above into Eq.~(\ref{tau1}), we obtain, in physical time units,
\begin{equation}\label{bad}
\tau=\frac{2\sqrt{\pi}}{\mu\sqrt{N}}\frac{R_0^{3/2}}{(R_0-1)^2 (R_0+1)^{1/2}}e^{N\Delta S}
\end{equation}
which coincides with the result of Ref.~\cite{kessler}.

\subsubsection{Examples 1-4 near the bifurcation point}
Because of their simplicity the examples 1-4, presented above, give identical results for the MTE near their corresponding bifurcation points, described by the equation $H_{qp}(0,0)=0$. The small distance to the bifurcation $\delta$ in all these examples is the ratio of the linear birth and death rates minus $1$:  $B-1$ in example 1,  $\lambda/\mu-1$ in example 2, and $R_0-1$ in examples 3 and 4. In all four examples $D_A=\sqrt{|H_{qqp}(0,0)|/H_{qpp}(0,0)}=1$. As there is only one linear branching process in each example, one has $K=1$, so Eq.~(\ref{A1bif1}) yields $\tilde{A}_1=-1$. As a result, in all four examples
\begin{equation}\label{tauuni}
\tau=\frac{\sqrt{2\pi}}{\alpha\sqrt{N}\delta^2}  \exp\left(\frac{N\delta^2}{2}\right)\,,
\end{equation}
where $\alpha$ denotes, in each example, the linear decay rate constant. To remind the reader, Eq.~(\ref{tauuni}) is
valid when $N\delta^2 \gg 1$ which, together with $\delta \ll 1$, yields the double inequality
$N^{-1/2} \ll \delta \ll 1$. At $\delta \sim  N^{-1/2}$ Eq.~(\ref{tauuni}) predicts the following
scaling relation for the MTE: $\tau \sim N^{1/2}/\alpha$, in agreement with Ref. \cite{Nasell}.

\section{Extinction Scenario B}\label{sB}
\subsection{General case: multi-step processes}
In this section we calculate the MTE and QSD for extinction scenario B.
Here extinction occurs along a trajectory composed of two segments: the non-zero-momentum
heteroclinic trajectory connecting the hyperbolic fixed points
$(n_2,0)$ and $(n_1,0)$ (the activation trajectory), and the zero-momentum segment going from $n=n_1$ to $n=0$ (the relaxation trajectory),
see Fig.~\ref{figB1a}.

A straightforward way to calculate the MTE starts with finding the  WKB solution for the QSD at $n\gg 1$. Then one should match it with the small-$n$ recursive solution (\ref{recsolution}), as in scenario A. The matching region in this case is $1\ll n\ll N$, or $N^{-1}\ll q\ll 1$. After having found the QSD, one can determine the MTE by using Eq.~(\ref{rec22}).

Actually, there is a shortcut to finding the MTE which does not require the knowledge of the small-$n$ recursive solution. This is because the solution includes a constant probability current flowing from a close vicinity of $n=n_1$ to a close vicinity of $n=0$, as shown below. This probability current is equal to the escape rate from the metastable state $q=q_2$, and it is determined by the WKB-asymptote of the QSD, with no use of the small-$n$ recursive solution. One of the objectives of our work, however, is to also find the QSD of the metastable state, and therefore we will follow the straightforward way.

In contrast to scenario A, where the WKB solution is determined solely by the fast mode, in scenario B the WKB solution is more complicated. Here the fast mode dominates to the right of the point $q=q_1$ (but not too close to $q_1$), whereas the slow-mode solution (\ref{slowmode}) dominates at $0<q<q_1$ (again, not too close to  $q=q_1$). Furthermore, the slow-mode solution diverges at $q=q_1$, and curing this divergence demands going beyond the WKB approximation in a boundary layer $|q-q_1|\ll 1$ where the fast and slow modes are strongly coupled. As a result, the QSD at $n\gg 1$ involves \textit{three} distinct asymptotes which need to be matched to one another. All this is very similar to what happens in other types of population escape problems: to an absorbing state at infinite population size \cite{MS} or to another metastable state \cite{EsK}. Much of the calculation is very similar to that of Refs. \cite{MS,EsK}, but we will present it here for completeness.

In the boundary layer $|q-q_1|\ll 1$  the momentum $p$ is small, that is fluctuations are weak. Here we can apply the van Kampen system size expansion \cite{kampen} to the quasi-stationary master equation (\ref{qsdmaster1}). Let us denote $f(q)=W_r(q) \pi(q) \simeq Nw_r(q) \pi(q)$ [it suffices to keep only the leading term in Eq.~(\ref{rateexp})]. Taylor-expanding $f(q-r/N)$ around $r=0$, we obtain
\begin{equation}\label{ftay}
f(q-r/N)\simeq f(q)-\frac{r}{N}f^{\prime}(q)+\frac{r^2}{2N^2}f^{\prime\prime}(q)\,.
\end{equation}
Plugging Eq.~(\ref{ftay}) into Eq.~(\ref{qsdmaster1}) and integrating once, we obtain
\begin{equation}
\sum_r -\frac{r}{N}f(q)+\frac{r^2}{2N^2}f^{\prime}(q)=\tilde{J}\,,
\end{equation}
where $\tilde{J} =const$. Now,
$$
f^{\prime}(q)=N[\pi^{\prime}(q)w_r(q)+\pi(q)w_r^{\prime}(q)]\,,
$$
but $\pi^{\prime}(q)\sim N\pi(q)$, so the second term in $f^{\prime}(q)$ is negligible. Therefore,
we obtain
\begin{equation}
-\pi(q)\sum_r r w_r(q)+\pi^{\prime}(q)\sum_r \frac{r^2}{2N}w_r(q)=\tilde{J}\,.
\end{equation}
The first term on the left corresponds to drift, the second one to diffusion. With the diffusion neglected, one obtains a (slow-mode) solution for $\pi(q)$ which diverges at fixed points of the  rate equation. The diffusion term cures this divergence by providing coupling between the slow mode and fast modes, as observed in Ref. \cite{MS}.

Now we use Eq.~(\ref{identities}) and evaluate the drift and diffusion terms in the vicinity of
$q=q_1$. In the drift term
$$
\sum_r rw_r(q)=H_p(q,0)\simeq
(q-q_1)H_{pq}(q_1,0)\,,
$$
while in the diffusion term it suffices to put $w_r(q)\simeq w_r(q_1)$. Denoting $x=(q-q_1)/l$, and $l^2=H_{pp}(q_1,0)/[NH_{pq}(q_1,0)]$  [as one can check, both $H_{pp}(q_1,0)$ and $H_{pq}(q_1,0)$ are positive], we obtain the boundary-layer equation \cite{MS}
\begin{equation}\label{fp}
\pi^{\prime}(x)-2x\pi(x)=J\,,
\end{equation}
where the rescaled constant current $J$ is to be found later. The general
solution of Eq.~(\ref{fp}) is
\begin{equation}\label{fpsolution}
\pi(x)=c_1 e^{x^2}+\frac{\pi J}{2} \mbox{erf}(x)\,.
\end{equation}
where $c_1$ is another constant.  Now we can match this solution to the slow-mode solution at $x<0$ and $|x|\gg 1$, that
is, at $N^{-1/2}\ll q_1-q \ll 1$. To eliminate the exponential
growth at $x<0$, one must choose $c_1=J\sqrt{\pi}/2$, so the
asymptote of the boundary-layer solution (\ref{fpsolution}) at $-x\gg 1$ becomes
\begin{equation}\label{fpleft}
\pi(x)\simeq -\frac{J}{2x}=\frac{J}{2(q_1-q)}\sqrt{\frac{H_{pp}(q_1,0)}{NH_{pp}(q_1,0)}}\,.
\end{equation}
The slow-mode solution (\ref{slowmode}) at $q\simeq q_1$ can be
approximated as
\begin{equation}\label{slowmodeasym}
\pi_s(q)= -\frac{A_s}{H_p(q,0)}\simeq
\frac{A_s}{(q_1-q)H_{pq}(q_1,0)}.
\end{equation}
Matching the two asymptotes, one obtains
\begin{equation}\label{J}
J=\frac{2A_s\sqrt{N}}{\sqrt{H_{pp}(q_1,0)H_{pq}(q_1,0)}}\,.
\end{equation}
To find the still unknown constant $A_s$, we have to match the $x\gg 1$ asymptote of the boundary-layer solution
(\ref{fpsolution}), which is
\begin{eqnarray}\label{fpright}
\pi(x)&\simeq&J\sqrt{\pi}e^{x^2}=\frac{2A_s\sqrt{\pi
N}}{\sqrt{H_{pp}(q_1,0)H_{pq}(q_1,0)}}\nonumber\\
&\times&\exp\left[\frac{NH_{pq}(q_1,0)}{H_{pp}(q_1,0)}(q-q_1)^2\right]\,,
\end{eqnarray}
with the  asymptote of the fast-mode solution at $N^{-1/2}\ll q-q_1
\ll 1$, which is
\begin{eqnarray}\label{fast2}
&&\hspace{-5mm}\pi(q)\simeq \sqrt{\frac{S^{\prime\prime}(q_2)}{2\pi N}}\nonumber\\
&&\hspace{-5mm}\times \;e^{N[S^{(f)}(q_2)-S^{(f)}(q_1)]+
S_1^{(f)}(q_2)-S_1^{(f)}(q_1)-(N/2)S^{\prime\prime}(q_1)(q-q_1)^2}.\nonumber\\
\end{eqnarray}
Here we have used the equalities $S^{\prime}(q_1)=p_a(q_1)=0$ and neglected
terms of order $q-q_1\ll 1$ in the exponent. Putting $q=q_1$ into Eq.~(\ref{hpq}), we obtain
\begin{equation}\label{paprime}
p_a^{\prime}(q_1)=S^{\prime\prime}(q_1)=-\frac{2H_{pq}(q_1,0)}{H_{pp}(q_1,0)},
\end{equation}
where $S^{\prime\prime}(q_1)<0$. Matching the asymptotes (\ref{fpright}) and (\ref{fast2}) and using Eq.~(\ref{paprime}), we find
\begin{eqnarray}\label{cs}
A_s&=&\frac{H_{pp}(q_1,0)\sqrt{|S^{\prime\prime}(q_1)|S^{\prime\prime}(q_2)}}{4\pi N}\nonumber\\
&\times& e^{N[S^{(f)}(q_2)-S^{(f)}(q_1)]+[S_1^{(f)}(q_2)-S_1^{(f)}(q_1)]}.
\end{eqnarray}
What is left is to find the MTE by matching the slow-mode solution at $q \ll 1$
with the recursive solution~(\ref{rec22}) at $q \gg N^{-1}$.
Using Eq.~(\ref{slowmode}) we obtain, at $q \ll 1$:
\begin{equation}
\pi_s(q)\simeq
-\frac{A_s}{qH_{pq}(0,0)}=\frac{A_s}{q|H_{pq}(0,0)|}\,,\label{slowa}
\end{equation}
where $H_{pq}(0,0)<0$ is given by Eq.~(\ref{hpqR}). Comparing this with Eq.~(\ref{rec22}) and using Eq.~(\ref{cs}), we obtain
\begin{eqnarray}
\tau&=&(\alpha N A_s)^{-1}=\frac{4\pi}{\alpha H_{pp}(q_1,0)\sqrt{|S^{\prime\prime}(q_1)|S^{\prime\prime}(q_2)}}\nonumber\\
&\times& e^{N[S^{(f)}(q_1)-S^{(f)}(q_2)]+[S_1^{(f)}(q_1)-S_1^{(f)}(q_2)]},\label{tau2}
\end{eqnarray}
where $\alpha$ in the linear decay rate constant in physical units \cite{physical}. The expression~(\ref{tau2}) for the MTE in scenario B is an important result of our work.  The leading term in the
exponent, proportional to $N$, is the effective entropy barrier to extinction. The proportionality factor is the absolute value of the area between the activation trajectory and relaxation trajectory \cite{Kamenev2}, see an example in Fig.~\ref{figB1a}. In contrast to scenario A, the pre-exponential factors in Eq.~(\ref{tau2}) are  $N$-independent.  Using Eqs.~(\ref{S1final}), (\ref{Psi}) and (\ref{paprime}), one can rewrite Eq.~(\ref{tau2}) in a more concise form:
\begin{eqnarray}
\tau=\frac{2\pi}{\alpha H_{pq}(q_1,0)}e^{N[S^{(f)}(q_1)-S^{(f)}(q_2)]+[\Psi(q_1)-\Psi(q_2)]}\,.\label{tau2simple}
\end{eqnarray}
As mentioned above, determining the MTE in scenario B does not require any information about the small-$n$ recursive solution. Furthermore, Eq.~(\ref{tau2}) formally coincides with the result of Escudero and Kamenev \cite{EsK}, who calculated a different quantity: the mean time to \textit{escape} from one metastable state into another. Finally, the same result (\ref{tau2}) can be also obtained for the mean time to escape to an absorbing state at infinity, as in the particular example considered by Meerson and Sasorov \cite{MS}. The reason for these coincidences is that, in all these systems, a constant probability current sets in beyond the repelling fixed point of the rate equation.   It is the magnitude of this current, carried by the slow WKB mode, rather than the exact nature of the target state for escape (an absorbing state at zero,  infinity or another metastable state), that determines,  in the leading and subleading orders in $N$, the mean escape rate from a metastable state.

To conclude this section, the QSD (another main result of this work) is given by  four overlapping asymptotes: (i) the recursive solution (\ref{recsolution}), valid for $1\leq n\ll N$, (ii) the slow-mode WKB solution (\ref{slowmode}), valid for $n_1-n\gg N^{1/2}$ and $n\gg 1$, (iii) the boundary-layer solution (\ref{fpsolution}), valid for $|n-n_1|\ll n_1$, and (iv) the fast-mode WKB solution (\ref{fastmode1}), valid for $n-n_1\gg N^{1/2}$.

\subsection{Single-step processes}
For completeness, we briefly consider the special case of single-step processes, where only $W_{\pm 1}(Nq)\equiv W_{\pm}(Nq)=Nw_{\pm}(q)+u_{\pm}(q)+\dots$ are present.
Here Eq.~(\ref{tau2}) simplifies considerably. Performing calculations similar to those in
scenario A (see Sec.~\ref{sA}) and using Eqs.~(\ref{identities}) and (\ref{S11}) and the fact that $w_+(q_{1,2})=w_-(q_{1,2})$, one obtains the MTE
\begin{eqnarray}
\tau=\frac{2\pi\,e^{\int_{q_1}^{q_2}\left(\frac{u_{+}}{w_{+}}-\frac{u_-}{w_-}\right)dq}}{\alpha w_+(q_2)\sqrt{|S^{\prime\prime}(q_1)|S^{\prime\prime}(q_2)}}
\;e^{N\int_{q_2}^{q_1}\ln \left( \frac{w_{-}}{w_+}\right)dq}.\label{tau21}
\end{eqnarray}
As expected, this result coincides with the single-step result of Ref.~\cite{EsK} for the mean time of a population switch between two metastable states.

\subsection{Extinction near saddle-node bifurcation point}
\begin{figure}
\includegraphics[width=2.6in,height=1.6in,clip=]{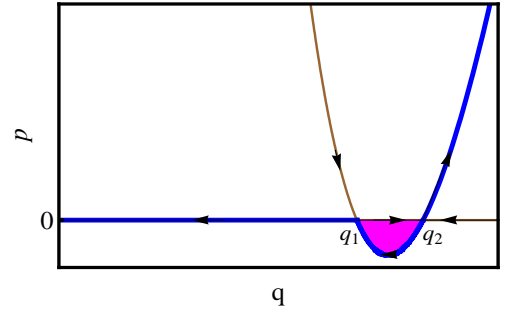}
\caption{(color online). Shown are typical zero-energy trajectories of the
WKB Hamiltonian (\ref{hamil}) in scenario B, close to the bifurcation where $q_2-q_1\ll 1$.
Here, the activation trajectory is a parabola, and $\Delta S$ given by Eq.~(\ref{sbifurB}) is the area of the shaded region.} \label{bifB}
\end{figure}

Here we calculate the MTE  near the characteristic (saddle-node) bifurcation of scenario B. At the bifurcation, the nontrivial attracting fixed point $q=q_2$  of the rate equation merges with the repelling point $q=q_1$. Above but near the bifurcation point $q_2-q_1\ll 1$. As a result, the momentum $p$ on the activation trajectory is much smaller than unity, see Fig.~\ref{bifB}. One can always define the parameter $N$ such that, at the bifurcation, $q_1=q_2= 1$. Furthermore, near the bifurcation $q_1=1-\delta$ and $q_2=1+\delta$, where the exact definition of $\delta\ll 1$ will appear shortly.

Let us Taylor-expand $H(q,p)$ from Eq.~(\ref{hamil}) in the vicinity of $q=1$ and $p=0$. As we expect $p_a(q)$ to be $\sim (q_2-q_1)^2$, we neglect  the terms of order $(q-1)p^2$ and higher
and arrive at the following equation for the zero-energy phase trajectories $p=p(q)$ close to $q=1$:
\begin{eqnarray}
H(q,p) &\simeq& p\sum_r \left[r w_r(1)+(q-1) \,r w_r^{\prime}(1)\right.\nonumber \\
&&\hspace{-8mm}+\left.\frac{1}{2}\,(q-1)^2 r w_r^{\prime\prime}(1)+\frac{1}{2}\, p r^2 w_r(1)\right]=0.\label{hbifB}
\end{eqnarray}
As can be checked \textit{a posteriori}, the terms in Eq.~(\ref{hbifB}) scale as follows: $H_p(1,0)\sim \delta^2$, $H_{qp}(1,0)\sim \delta^2$, $H_{qqp}(1,0)= {\cal O}(1)$, and $H_{pp}(1,0)={\cal O}(1)$. Therefore, the term $(q-1)H_{qp}(1,0)\sim \delta^{3}$ can be neglected.
The nontrivial solution of Eq.~(\ref{hbifB}) yields the activation trajectory $p=p_a(q)$: a parabola with the roots $q=q_1$ and $q=q_2$.  To simplify the notation, we use Eq.~(\ref{identities}) and evaluate the small difference $q_2-q_1$ by expanding the algebraic equation $H_p(q,0)=0$ in the vicinity of $q=1$.  Neglecting the term $(q-1)H_{qp}(1,0)$, we obtain
\begin{equation}\label{qbifB}
\delta\equiv \frac{q_2-q_1}{2}=\sqrt{\frac{2H_p(1,0)}{|H_{qqp}(1,0)|}}\,,
\end{equation}
where $H_{qqp}(q_1,0)<0$. The activation trajectory can be written as
\begin{equation}\label{actbif1B}
p_a(q)=\frac{|H_{qqp}(1,0)|}{H_{pp}(1,0)}(q-q_1)(q-q_2)\,.
\end{equation}
As $S^{\prime\prime}(q)=p_a^{\prime}(q)$, we find $S^{\prime\prime}(q_2)=-S^{\prime\prime}(q_1)=2\delta^2 |H_{qqp}(1,0)|/H_{pp}(1,0)$.
Furthermore, the action $S^{(f)}(q)=\int^q p_a(\xi)d\xi$ is given by
\begin{eqnarray}\label{actionbifB}
S^{(f)}(q)\!=\!\frac{|H_{qqp}(1,0)|}{H_{pp}(1,0)}\left[\frac{q^3}{3}\!-\!\frac{q^2}{2}(q_1\!+\!q_2)\!+\!q q_1 q_2\right],
\end{eqnarray}
whereas
\begin{equation}\label{sbifurB}
\Delta S=S^{(f)}(q_1)-S^{(f)}(q_2)=\frac{4|H_{qqp}(1,0)|}{3H_{pp}(1,0)}\delta^{3}
\end{equation}
is the area of the shaded region in Fig.~\ref{bifB}.

As in scenario A, the sub-leading WKB correction vanishes near the bifurcation. Indeed, we Taylor-expand Eq.~(\ref{s1calculation}) in the vicinity of $q=1$ and $p=0$, keep only leading-order terms, and obtain
\begin{equation}
\sum_r r w_r(1)S_1^{\prime}(q)-\frac{1}{2}r^2 w_r(1)(q-1) p_a^{\prime\prime}(1)-rw_r^{\prime\prime}(1)\simeq 0\,.
\end{equation}
Using Eq.~(\ref{actbif1B}), we find that the second and third terms cancel out, and so $S_1^{(f)}(q)$ can be chosen zero. As a result, Eq.~(\ref{tau2}) yields the MTE near the bifurcation:
\begin{equation}\label{taubifB}
\tau=\frac{2\pi}{\alpha |H_{qqp}(1,0)|\,\delta}\,\exp\left(\frac{4}{3}N D_B^2\delta^{3}\right),
\end{equation}
where $H_{qqp}(1,0)$ and $D_B=\sqrt{|H_{qqp}(1,0)|/H_{pp}(1,0)}={\cal O}(1)$ should be evaluated \textit{at} the bifurcation. The applicability criterion of this result is $N\delta^{3}\gg 1$. For sufficiently large $N$ this strong inequality is compatible with the strong inequality  $\delta \ll 1$ which describes closeness to the bifurcation. At $\delta \sim N^{-1/3}$ Eq.~(\ref{taubifB}) predicts the following scaling of the MTE with $N$: $\tau \sim N^{1/3}/\alpha$.

We notice that Eq.~(\ref{taubifB}) does not require any information about the QSD in the region of small $n$. Indeed, as was mentioned in section V A,  the exact nature of the target state is of no significance here. Note that the same  scaling  of the effective entropy barrier with the distance from the bifurcation  appears in the context of escape from one metastable state to another \cite{dykman}.  Finally, the same scaling near the bifurcation is observed in \textit{continuous} systems, driven by external delta-correlated gaussian noise and therefore describable by a Fokker-Planck equation \cite{dykman80,3/2}.

\subsection{Example: reactions $2A \rightleftarrows 3A$ and $A\to 0$}
Let us illustrate the extinction scenario B on the following set of
reactions: binary
reproduction $2A\stackrel{\lambda}{\rightarrow} 3A$, the reverse
process $3A\stackrel{\sigma}{\rightarrow} 2A$, and  linear decay $A\stackrel{\mu}{\rightarrow} \emptyset$. Here
\begin{eqnarray}
\hspace{-4mm}W_1\!=\!\frac{\lambda n(n-1)}{2}\,,\;W_{-1}=\mu n +\frac{\sigma n(n-1)(n-2)}{6}.\label{rates}
\end{eqnarray}
Rescaling time $\mu t \to t$, and denoting $\gamma=1-\delta^2$, $\delta^2=1-8
\sigma \mu/(3 \lambda^2)>0$, and $N=3\lambda/(2 \sigma)$, we arrive at Eq.~(\ref{rateexp}) with
\begin{eqnarray}
w_{-1}&=&\frac{q^3}{\gamma}+q\;,\;\;\;w_1=\frac{2q^2}{\gamma}\;,\nonumber\\
u_{-1}&=&-\frac{3q^2}{\gamma}\;,\;\;\;u_1=-\frac{2q}{\gamma}.\label{rates2}
\end{eqnarray}
In the rescaled notation the fixed points are $q=0$ (attracting point), $q_1=1-\delta$ (repelling point), and $q_2=1+\delta$: another attracting point around which the metastable population resides.
The WKB Hamiltonian (\ref{hamil}) takes the form
\begin{equation}\label{hamilex2}
H(q,p)=\left(\frac{q^3}{\gamma}+q\right)(e^{-p}-1)+\frac{2q^2}{\gamma}(e^p-1)\,.
\end{equation}
Solving the equation $H[q,p_a(q)]=0$ yields
\begin{equation}\label{zeroE}
p_a(q)=S^{\prime}(q)=\ln\left(\frac{q}{2}+\frac{\gamma}{2q}\right)\,.
\end{equation}
Therefore,
\begin{equation}
S^{\prime\prime}(q)=\frac{q^2-\gamma}{q^2(q+\gamma)}\,,
\end{equation}
$S^{\prime\prime}(q_1)=-\delta/(1-\delta)$, and
$S^{\prime\prime}(q_2)=\delta/(1+\delta)$.
Furthermore, using Eq.~(\ref{rates2}) we obtain
\begin{eqnarray}\label{deltaS1}
\hspace{-3mm}\int_{q_2}^{q_1}\ln  \frac{w_{-}}{w_+} \,dq=2\left(\delta-\sqrt{1-\delta^2} \arctan
\frac{\delta}{\sqrt{1-\delta^2}}\right)
\end{eqnarray}
and
$$\exp\left[-\int_{q_2}^{q_1}\left(\frac{u_{+}}{w_{+}}-\frac{u_-}{w_-}\right)dq\right]=
\sqrt{\frac{1+\delta}{1-\delta}}.$$
Plugging everything into Eq.~(\ref{tau21}) yields the MTE in physical units:
\begin{equation}\label{mte2}
\tau=\frac{\pi(1-\delta)}{\mu\delta}\,e^{N \Delta S}\,.
\end{equation}
Here
\begin{equation}
\Delta S = 2 \left(\delta-\sqrt{1-\delta^2}
\arctan \frac{\delta}{\sqrt{1-\delta^2}}\right)
\label{deltaS2}
\end{equation}
is a monotone increasing function of $\delta$; its asymptotes are
$$
\Delta S=\left\{\begin{array}{ll}
(2/3)\delta^3 + (4/15) \delta^5+\dots,\;\;\;\delta\ll 1\,, \\
 2-\pi\sqrt{2 (1-\delta)}+\dots,\;\;\;1-\delta\ll 1\,.
\end{array}
\right.
$$
Near the bifurcation, $\delta\ll 1$, Eq.~(\ref{mte2}) becomes
\begin{equation}\label{mte2bif}
\tau=\frac{\pi}{\alpha\delta}\exp\left(\frac{2}{3}N \delta^3\right)\,.
\end{equation}
This is in agreement with Eq.~(\ref{taubifB}). Indeed, at the bifurcation one has $H_{pp}(1,0)= 4$,
$H_{qqp}(1,0)= -2$, and so $D_B^2= 1/2$. Equation~(\ref{mte2bif}) is
valid when $N\delta^3 \gg 1$. This inequality, combined with $\delta \ll 1$, yields the double inequality
$N^{-1/3} \ll \delta \ll 1$.

\begin{figure}
\includegraphics[width=6.4cm,height=5.0cm,clip=]{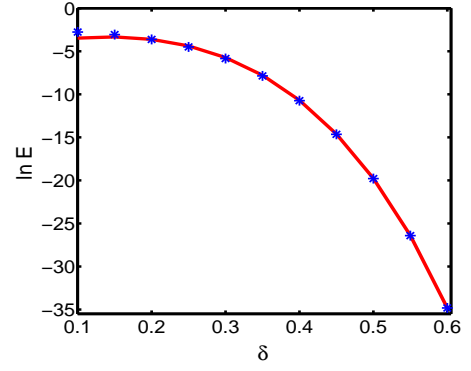}
\caption{(color online). The natural logarithm of the extinction rate $E=1/\tau$ versus $\delta$ for $N=200$ for the reactions $A\stackrel{\mu}{\rightarrow} \emptyset$, $2A\stackrel{\lambda}{\rightarrow} 3A$ and  $3A\stackrel{\sigma}{\rightarrow} 2A$. The analytical
result (\ref{mte2}) (the solid line) is compared with the quantity $\,-[\ln(1-P_0^{num}(t))]/t$ (the asterisks) extracted from a numerical solution  of (a truncated) master equation
Eq.~(\ref{master}) with rates (\ref{rates}).} \label{e1vsdelta}
\end{figure}

We compared our analytical result~(\ref{mte2}) with numerical solutions of (a truncated)  master equation (\ref{master}) with rates (\ref{rates}) at $N=200$ and different values of $\delta$. The comparison is presented in Fig.~\ref{e1vsdelta}. Very good agreement is observed for not too small $\delta$, when the effective entropy barrier $N \Delta S$ is sufficiently high.

\begin{figure}
\includegraphics[width=6.7cm,height=4.8cm,clip=]{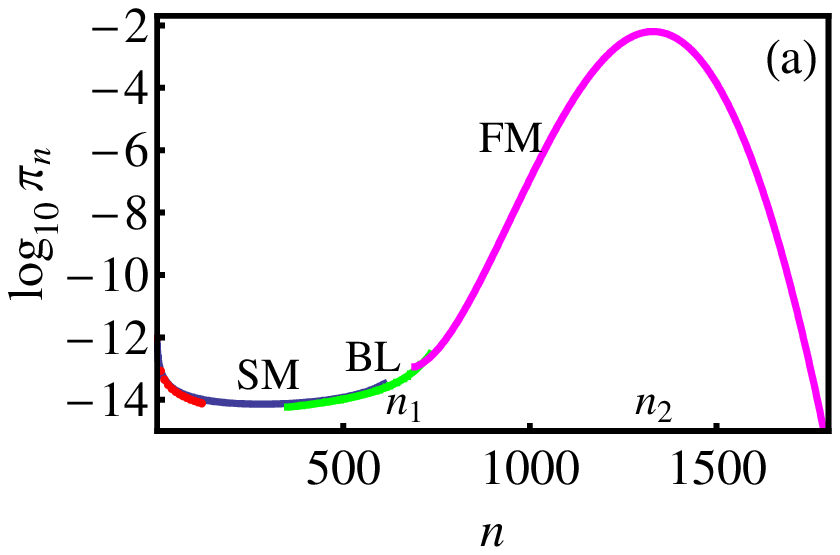}
\hspace{5mm}
\includegraphics[width=6.7cm,height=4.8cm,clip=]{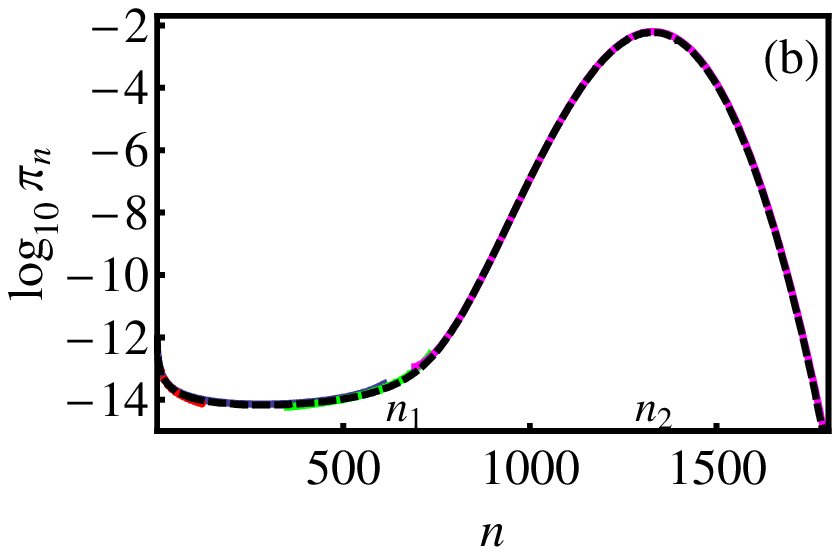}
\caption{(color online). (a): the QSD $\pi_n$ for $\delta=1/3$ and
$N=10^3$ for the reactions $A\stackrel{\mu}{\rightarrow} \emptyset$, $2A\stackrel{\lambda}{\rightarrow} 3A$ and  $3A\stackrel{\sigma}{\rightarrow} 2A$. The QSD includes four overlapping asymptotes: the
fast-mode solution (FM), the slow-mode solution (SM), the boundary-layer solution (BL), and the
small-$n$ asymptote. (b) a comparison
between the QSD in (a) (the solid line) and a numerical solution of
(a truncated) master equation (\ref{master}) with rates (\ref{rates}) (the dashed
line).}\label{qsd2}
\end{figure}

To determine the QSD in this example, one needs to find $S^{(f)}(q)$ and $S_1^{(f)}(q)$ from Eqs.~(\ref{sfast}) and (\ref{s1fast}):
\begin{eqnarray}\label{Sexamp2}
S^{(f)}(q)&=&2\sqrt{\gamma}\arctan\left(\frac{q}{\sqrt{\gamma}}\right)+q\left[\ln\left(\frac{\gamma+q^2}{2q}\right)-1\right]\nonumber\\
S_1^{(f)}(q)&=&-\ln\left(\frac{\gamma+q^2}{q^{5/2}}\right)\,.
\end{eqnarray}
Then Eqs.~(\ref{fastmode1}), (\ref{slowmode}), (\ref{recsolution}) and (\ref{fpsolution}) yield the QSD in terms of four overlapping asymptotes. The QSD and its comparison with  numerics are shown in Fig.~\ref{qsd2}, and very good agreement is observed.

\section{Summary}\label{conclusion}
This work dealt with extinction of an isolated long-lived stochastic population describable by a continuous-time Markov process. We have identified two generic extinction scenarios, A and B, based on the stability properties of the fixed points of the population size dynamics, predicted by the  rate equation. For each of the two scenarios we have calculated the mean time to extinction (MTE) and the quasi-stationary probability distribution (QSD).  The calculations involve a systematic use of WKB method, where $1/N$,  the typical inverse population size in the metastable state, serves as a small parameter of the theory. The WKB theory is supplemented by two additional approximations in the regions where it breaks down. One of them is a small-$n$ expansion of the quasi-stationary master equation which brings about a recursion relation (in both scenarios A and B). The second one is the Fokker-Planck equation, obtained via the van Kampen system-size expansion. The latter is valid in small regions around the non-trivial fixed points of the rate equation (in scenario B).

The theory is not limited to single-step stochastic processes, although for such processes our general results simplify considerably. The results also simplify near the characteristic bifurcations of scenarios A and B. A number of previous results for the mean time to extinction follow from our equations in particular cases.

We have observed that, in models belonging to extinction scenario B, the mean time to population extinction formally coincides, in the leading and subleading orders in $N$, with the mean time to population \textit{escape} in stochastic population models where the metastable population switches  to another metastable state or to an absorbing state at infinity. The reason for this coincidence is that, in all these systems, a constant probability current sets in beyond the repelling fixed point of the rate equation. It is the magnitude of this current, rather than the precise nature of the target state for escape, which determines,  in the leading and subleading orders in $N$, the mean escape rate from a metastable state.

The situation is quite different in extinction scenario A. Here the exact nature of the target state (the absorbing state at $n=0$) or, more precisely, the rate constants of the effective linear branching processes at small $n$, affects the pre-exponent of the mean time to extinction.

\subsection*{Acknowledgments}
We are very grateful to B. Derrida, M. Dykman, A. Kamenev, and P.V. Sasorov for discussions. B.~M. and M.~A. were supported by the Israel Science
Foundation (Grant No. 408/08). M.~A. was supported by the Clore Foundation.

\section*{Appendix A}
\renewcommand{\theequation}{A\arabic{equation}}
\setcounter{equation}{0}
Here we derive Eq.~(\ref{cires}) for the arbitrary constants $C_i$ appearing in Eq.~(\ref{fn}). For that we need to solve the following set of  $K+1$ linear equations:
\begin{equation}\label{eqset}
    C_0+\sum_{i=1}^K  \lambda_i^{-m} C_i =f_m\,,\;\;\;m=1,2, \dots, K+1\,,
\end{equation}
where the constants $f_2,f_3,\dots,f_{K+1}$ can be expressed via $f_1$ by using the recursive
Eq.~(\ref{QSDR3}) with $n=1,2,\dots,K$ (assuming that $f_j=0$ for $j<1$).

As before, we denote the roots of Eq.~(\ref{pol2}) by
$\lambda_0,\lambda_1,\dots,\lambda_K$, where $\lambda_0=1$, and define the following quantities:
\begin{eqnarray}
s_0^{(0)}&=&1\nonumber\\
s_1^{(0)}&=&\lambda_1+\lambda_2+\lambda_3+\cdots+\lambda_K\nonumber\\
s_2^{(0)}&=&\lambda_1\lambda_2+\lambda_2\lambda_3+\lambda_3\lambda_4+\cdots+\lambda_{K-1}\lambda_{K}\nonumber\\
s_3^{(0)}&=&\lambda_1\lambda_2\lambda_3+\lambda_1\lambda_3\lambda_4+\lambda_1\lambda_2\lambda_4+\cdots\nonumber\\
&&\cdots\cdots\nonumber\\
s_K^{(0)}&=&\lambda_1\lambda_2\cdots\lambda_K\,.\label{s00}
\end{eqnarray}
The superscript $(0)$ means that the $K$ roots which
contribute to $s_i^{(0)}$ are all of the roots of Eq.~(\ref{pol2}) \textit{except} $\lambda_0$. In the same manner we can define for
$1\leq i\leq K$
\begin{eqnarray}
s_0^{(i)}&=&1\nonumber\\
s_1^{(i)}&=&1+\lambda_1+\lambda_2+\cdots+\lambda_{i-1}+\lambda_{i+1}+\cdots+\lambda_K\nonumber\\
s_2^{(i)}&=&\sum_{\begin{array}{c}0\leq j<m\leq K\\j,m\neq i\\\end{array}}^K\hspace{-10mm}\lambda_j\lambda_m\nonumber\\
&&\cdots\cdots\nonumber\\
s_K^{(i)}&=&\lambda_1\lambda_2\cdots\lambda_{i-1}\lambda_{i+1}\cdots\lambda_K\,,\label{si0}
\end{eqnarray}
where the root $\lambda_i$ does \textit{not} contribute.

To obtain the coefficient $C_0$ we
multiply the first of Eqs.~(\ref{eqset}) by
$s_0^{(0)}$, the second by $-s_1^{(0)}$,..., and finally the last by $(-1)^K s_K^{(0)}$. By adding the equations we obtain
\begin{eqnarray}\label{c0calc1}
&&\hspace{-4mm}f_1-s_1^{(0)}f_2+s_2^{(0)}f_3-\cdots+(-1)^K s_K^{(0)}f_{K+1}\nonumber\\
&&\hspace{-4mm}=C_0\left[1-s_1^{(0)}+s_2^{(0)}-\cdots+(-1)^K s_K^{(0)}\right]\nonumber\\
&&\hspace{-4mm}+C_1\left[\lambda_1^{-1}-\lambda_1^{-2}s_1^{(0)}+\lambda_1^{-3}s_2^{(0)}-\cdots+(-1)^K \lambda_1^{-K-1}s_K^{(0)}\right]\nonumber\\
&&\hspace{-4mm}+\cdots+C_K\left[\lambda_K^{-1}-\lambda_K^{-2}s_1^{(0)}+\cdots+(-1)^K \lambda_K^{-K-1}s_K^{(0)}\right].\nonumber\\
\end{eqnarray}
By using Eqs.~(\ref{s00}) we can rewrite the coefficient of $C_0$ in the right hand side of Eq.~(\ref{c0calc1}) as
\begin{equation}\label{c01}
1-s_1^{(0)}+s_2^{(0)}-\cdots+(-1)^K s_K^{(0)}=(1-\lambda_1)(1-\lambda_2)\cdots(1-\lambda_K).
\end{equation}
Furthermore, the coefficient of $C_i$ in Eq.~(\ref{c0calc1}) satisfies
\begin{eqnarray}
&&\hspace{-12mm}\lambda_i^{-1}-\lambda_i^{-2}s_1^{(0)}+\lambda_i^{-3}s_2^{(0)}-\cdots+(-1)^K\lambda_i^{-K-1}s_K^{(0)}\nonumber\\
&&\hspace{-12mm}=\frac{(\lambda_i-\lambda_1)(\lambda_i-\lambda_2)\cdots(\lambda_i-\lambda_j)\cdots(\lambda_i-\lambda_K)}{\lambda_i^{K+1}}.
\end{eqnarray}
Clearly, this expression vanishes for all $i\geq 1$.
Therefore,  Eqs.~(\ref{c0calc1}) and (\ref{c01}) yield
\begin{equation}\label{c02}
C_0=\frac{f_1-s_1^{(0)}f_2+s_2^{(0)}f_3-\cdots+(-1)^K s_K^{(0)}f_{K+1}}{\displaystyle\prod_{i=1}^{K}(1-\lambda_i)}
\end{equation}

To calculate the numerator of Eq.~(\ref{c02}) we use Viete's formula. Given a polynomial
equation
$$a_K x^K + \cdots + a_1 x + a_0=0\,,$$ whose roots are
$\lambda_1,\lambda_2,\cdots,\lambda_K$, the expressions $s_i^{(0)}$, given by Eq.~(\ref{s00}), satisfy
\begin{equation}\label{siviete}
s_i^{(0)}=(-1)^i\frac{a_{K-i}}{a_K}\,.
\end{equation}
Let us apply this formula to Eq.~(\ref{pol3}) which has exactly the
roots $\lambda_1,\lambda_2,\cdots,\lambda_K$. For convenience, we rewrite Eq.~(\ref{pol3}) as
\begin{equation}\label{pol4}
1-\lambda\sum_{r=1}^K w_{r}^{\prime}(0)-\lambda^2\sum_{r=2}^K w_{r}^{\prime}(0)-\cdots-\lambda^K w_{K}^{\prime}(0)=0\,.
\end{equation}
In terms of the coefficients $a_i$ we have $a_0=1$, $a_1=-[w_{1}^{\prime}(0)+\cdots+w_{K}^{\prime}(0)]$,
$a_2=-[w_{2}^{\prime}(0)+\cdots+w_{K}^{\prime}(0)]\,,\cdots$, and $a_K=-w_{K}^{\prime}(0)$. Therefore, using Eqs.~(\ref{c02}) and (\ref{siviete}), we can rewrite $C_0$ as
\begin{eqnarray}
C_0=
\frac{f_{K+1}-f_{K} [w_{1}^{\prime}(0)+\cdots+w_{K}^{\prime}(0)]-\cdots-f_1 w_{K}^{\prime}(0)}
{a_K\displaystyle\prod_{i=1}^{K}(1-\lambda_i)}.\nonumber\\\label{C}
\end{eqnarray}
Now, using Eq.~(\ref{QSDR3}) with $n=K$, we obtain a relation
between $f_{K+1}$ and $f_{j\leq K}$. Plugging it into (\ref{C}) we have
\begin{eqnarray}
C_0&=&\frac{1}{a_K\displaystyle\prod_{i=1}^{K}(1-\lambda_i)}\left\{f_{K}-f_{K-1} [w_{1}^{\prime}(0)+\cdots+w_{K}^{\prime}(0)]\right.\nonumber\\
&-&\left.\cdots-f_1 [w_{K-1}^{\prime}(0)+w_{K}^{\prime}(0)]\right\}.\label{CC}
\end{eqnarray}
One can use this argument repeatedly $K-1$ more times, and obtain that the expression in the curly brackets in Eq.~(\ref{CC}) equals $f_1$. In addition, by virtue of Eq.~(\ref{siviete}) $w_{K}^{\prime}(0)$ satisfies
\begin{equation}\label{rk}
w_{K}^{\prime}(0)=-a_K=\frac{(-1)^{K+1}}{\lambda_1\lambda_2\cdots\lambda_K}\,.
\end{equation}
Therefore, $C_0$ is given by
\begin{equation}\label{c0}
C_0=\frac{(-1)^K
f_1\displaystyle\prod_{i=1}^{K}\lambda_i}{\displaystyle\prod_{i=1}^{K}(1-\lambda_i)}=\frac{f_1}{\displaystyle\prod_{i=1}^{K}(1-\lambda_i^{-1})}\,,
\end{equation}
thereby proving Eq.~(\ref{cires}) for $i=0$.

This proof can be generalized to the rest of the coefficients $C_i$. Here one has to multiply
the first of Eqs.~(\ref{eqset}) by
$s_0^{(i)}$, the second by $-s_1^{(i)}$,..., and finally the last by $(-1)^K s_K^{(i)}$, and add all the equations.
This yields
\begin{eqnarray}\label{cicalc1}
&&\hspace{-4mm}f_1-s_1^{(i)}f_2+s_2^{(i)}f_3-\cdots+(-1)^K s_K^{(i)}f_{K+1}\nonumber\\
&&\hspace{-4mm}=C_0\left[1-s_1^{(i)}+s_2^{(i)}-\cdots+(-1)^K s_K^{(i)}\right]\nonumber\\
&&\hspace{-4mm}+C_1\left[\lambda_1^{-1}-\lambda_1^{-2}s_1^{(i)}+\lambda_1^{-3}s_2^{(i)}-\cdots+(-1)^K \lambda_1^{-K-1}s_K^{(i)}\right]\nonumber\\
&&\hspace{-4mm}+\cdots+C_K\left[\lambda_K^{-1}-\lambda_K^{-2}s_1^{(i)}+\cdots+(-1)^K \lambda_K^{-K-1}s_K^{(i)}\right].\nonumber\\
\end{eqnarray}
The coefficient of $C_i$ in the right hand side of Eq.~(\ref{cicalc1}) satisfies
\begin{eqnarray}
&&\hspace{-4mm}\lambda_i^{-1}-\lambda_i^{-2}s_1^{(i)}+\lambda_i^{-3}s_2^{(i)}-\cdots+(-1)^K\lambda_i^{-K-1}s_K^{(i)}\nonumber\\
&&\hspace{-4mm}=\frac{(\lambda_i\!-\!1)(\lambda_i\!-\!\lambda_1)\cdots(\lambda_i\!-\!\lambda_{i-1})(\lambda_i\!-\!\lambda_{i+1})\cdots(\lambda_i\!-\!\lambda_K)}{\lambda_i^{K+1}},\nonumber\\
\end{eqnarray}
because $\lambda_i$ is absent from $s_j^{(i)}$, see Eq.~(\ref{si0}). On the other hand, the coefficients of all other $C_{j\neq i}$ in the right hand side of Eq.~(\ref{cicalc1}) can be shown to be equal zero. Therefore,
\begin{equation}\label{ci2}
C_i=\frac{\lambda_i^{K+1}\left[f_1-s_1^{(i)}f_2+s_2^{(i)}f_3-\cdots+(-1)^K s_K^{(i)}f_{K+1}\right]}{\displaystyle\prod_{\begin{array}{c}j=0\\j\neq
i\\\end{array}}^{K}(\lambda_i-\lambda_j)}
\end{equation}
By using the recursive equation (\ref{QSDR3}) repeatedly and by using Eq.~(\ref{si0}), one finally obtains
Eq.~(\ref{cires}).

Finally, $C_0$ from Eq.~(\ref{c0}) can be
expressed through the reaction rate constants $w_{1}^{\prime}(0),w_{2}^{\prime}(0),\cdots,w_{K}^{\prime}(0)$, see Eq.~(\ref{c00main}).
Indeed, let us expand the denominator of
Eq.~(\ref{c0}) and use Eq.~(\ref{s00}):
\begin{eqnarray}\label{calc}
&&(1-\lambda_1^{-1})(1-\lambda_2^{-1})\cdots(1-\lambda_K^{-1})=1-\sum_{i}\lambda_{i}^{-1}\nonumber\\
&+&\sum_{i>j}\lambda_i^{-1}\lambda_j^{-1}+\cdots+(-1)^K
\lambda_1^{-1}\cdots\lambda_{K}^{-1}\nonumber\\
&=&\frac{(-1)^K}{\lambda_1\cdots\lambda_K}\left[(-1)^K
s_K^{(0)}+(-1)^{K-1}s_{K-1}^{(0)}+\cdots+1\right].\nonumber\\
\end{eqnarray}
Using Eqs.~(\ref{siviete}) and (\ref{rk}), we rewrite Eq.~(\ref{calc}) as
\begin{eqnarray}\label{calc1}
&&(1-\lambda_1^{-1})(1-\lambda_2^{-1})\cdots(1-\lambda_K^{-1})=1-[w_{1}^{\prime}(0)+\cdots\nonumber\\
&&+w_{K}^{\prime}(0)]-[w_{2}^{\prime}(0)+\cdots+w_{K}^{\prime}(0)]-\cdots-w_{K}^{\prime}(0)\nonumber\\
&&\hspace{-7mm}=1-w_{1}^{\prime}(0)-2w_{2}^{\prime}(0)-3w_{3}^{\prime}(0)-\cdots-Kw_{K}^{\prime}(0)\,.
\end{eqnarray}
Plugging this into Eq.~(\ref{c0}) one obtains Eq.~(\ref{c00main}).

\section*{Appendix B}
\renewcommand{\theequation}{B\arabic{equation}}
\setcounter{equation}{0}
Here we show that, in extinction scenario A, the two real and positive roots of Eq.~(\ref{pol2}) are $\lambda_0=1$ and $0<\lambda_1<1$, whereas all other roots obey the inequality $|\lambda_{i>1}|>\lambda_1$. We start by showing that the positive root of Eq.~(\ref{pol3}) obeys the inequalities $0<\lambda_1<1$. The left hand side of Eq.~(\ref{pol3})
side is a monotone decreasing function of $\lambda$. At $\lambda=1$ it is equal to
$1-w_{1}^{\prime}(0)-2w_{2}^{\prime}(0)-\cdots-Kw_{K}^{\prime}(0)$. This quantity is negative, as $n=0$ is a repelling
fixed point, so the rescaled reaction rate constants satisfy the inequality $w_{1}^{\prime}(0)+2w_{2}^{\prime}(0)+\cdots+Kw_{K}^{\prime}(0)-1>0$.
On the other hand, at $\lambda=0$ the left hand side is $1$. Hence, $0<\lambda_1<1$.

Now we will prove by contradiction that all other (negative or complex) roots
satisfy the inequaltity
$|\lambda_{i>1}|>\lambda_1$. Assume by contradiction that there
exists a root $\lambda_j$ so that $|\lambda_j|\leq \lambda_1$. Denote $\lambda_j=a
e^{i\theta}$. Then by assumption $a\leq \lambda_1$. Substituting $\lambda_j$
into Eq.~(\ref{pol3}) we have
\begin{eqnarray}
&&1-w_{1}^{\prime}(0)\lambda_j-\cdots-w_{K}^{\prime}(0)(\lambda_j+\cdots+\lambda_j^{K})\nonumber\\
&&\hspace{-5mm}=1-w_{1}^{\prime}(0) a\cos\theta-\cdots\nonumber\\
&&\hspace{-5mm}-w_{K}^{\prime}(0)(a\cos\theta+\cdots+a^K\cos K\theta)+i\,(\cdots)=0,\label{pr1}
\end{eqnarray}
where both real and imaginary parts have to vanish separately. Now we
substitute $\lambda_1$ into Eq.~(\ref{pol3}) and use Eq.~(\ref{pr1})
\begin{eqnarray}\label{pr2}
0&=&1-w_{1}^{\prime}(0)\lambda_1-\cdots-w_{K}^{\prime}(0)(\lambda_1+\cdots+\lambda_1^{K})\nonumber\\
&\leq&1-w_{1}^{\prime}(0) a-\cdots-w_{K}^{\prime}(0)(a+\cdots+a^K)\nonumber\\
&<&1-w_{1}^{\prime}(0)a\cos\theta-\cdots\nonumber\\
&-&w_{K}^{\prime}(0)(a\cos\theta+\cdots+a^K\cos K\theta)=0,
\end{eqnarray}
where the last inequality holds since $\lambda_j$ is complex or
negative, so $\theta\neq 0$, and there exists some $m$ for which $\cos
m\theta<1$.

Eq.~(\ref{pr2}) shows a contradiction $0<0$, so all roots obey
$|\lambda_{i>1}|>\lambda_1$. As a result, at $n\gg 1$, the recursive
solution (\ref{recsolution}) reduces to Eq.~(\ref{rec1}), where $A_1$ is given by Eq.~(\ref{c1}).

Finally, we show that $A_1>0$, and so the asymptote (\ref{rec1}) is always positive. First, by using Eq.~(\ref{rk}), one can see that the numerator in Eq.~(\ref{c1}) is always negative. What is the sign of the denominator? Here $\lambda_1-\lambda_0<0$, whereas all other terms in the product are positive. Indeed, for $\lambda_j<0$ one has $\lambda_1-\lambda_j>0$. For any complex $\lambda_j$, there is also a complex conjugate root $\lambda_k=\overline{\lambda_j}$. Therefore, by writing $\lambda_j=a+ i b$, one has $(\lambda_1-\lambda_j)(\lambda_1-\overline{\lambda_j})=(\lambda_1-a)^2+b^2>0$. So, $A_1$ is always positive, and so is the asymptote $\pi(q)$ from Eq.~(\ref{rec1}).

\section*{Appendix C}
\renewcommand{\theequation}{C\arabic{equation}}
\setcounter{equation}{0}
Here we show that, in extinction scenario B, the root $\lambda_0=1$ of Eq.~(\ref{pol2}) has the smallest absolute value among all of the roots.
To this end we will prove that the roots of Eq.~(\ref{pol3}) obey the inequality $|\lambda_{j>0}|>1$.
Let us denote $\lambda_{j>0}=a_j
e^{i\theta_j}$, in general a complex number, and assume by contradiction that $a_j\leq 1$. Plugging $\lambda=\lambda_j$ into Eq.~(\ref{pol3}),
we obtain
\begin{eqnarray}\label{pol6}
&&\hspace{-4mm}1-a_j \cos(\theta_j)w_{1}^{\prime}(0)-[a_j \cos(\theta_j)\!+\!a_j^2 \cos(2\theta_j)]w_{2}^{\prime}(0)-\cdots\nonumber\\
&&\hspace{-4mm}-[a_j \cos(\theta_j)+\cdots+a_j^{K}\cos(K\theta_j)]w_{K}^{\prime}(0)+i\,(\cdots)=0,\nonumber\\
\end{eqnarray}
where the real and imaginary parts must vanish separately. As we have assumed $a_j\leq 1$, we have
$a_j^{m}\cos(m\theta_j)\leq 1$ for all integer $m\geq 0$. Therefore, we can write for the real part of Eq.~(\ref{pol6}):
\begin{eqnarray}\label{pol7}
&&\hspace{-7mm}0=1-a_j \cos(\theta_j)w_{1}^{\prime}(0)-[a_j \cos(\theta_j)+a_j^2 \cos(2\theta_j)]w_{2}^{\prime}(0)\nonumber\\
&-&\cdots- [a_j \cos(\theta_j)+\cdots+a_j^{K}
\cos(K\theta_j)]w_{K}^{\prime}(0)\nonumber\\&\geq& 1-w_{1}^{\prime}(0)-2w_{2}^{\prime}(0)-\cdots-Kw_{K}^{\prime}(0)
>0\,,
\end{eqnarray}
where the last inequality follows from $n=0$ being an attracting fixed point of the rate equation.
Equation~(\ref{pol7}) shows a contradiction $0>0$. Hence $a_j>1$, and all the roots of Eq.~(\ref{pol3})
obey the inequality $|\lambda_{j>0}|>1$.


\begin{thebibliography}{99}
\bibitem{bartlett} M.S. Bartlett, \textit{Stochastic Population Models in Ecology
and Epidemiology} (Wiley, New York, 1961).
\bibitem{assessment} S. R. Beissinger and D. R. McCullough (Editors),
\textit{Population Viability Analysis} (University of Chicago Press, Chicago, 2002).
\bibitem{epidemic} H. Andersson and T. Britton,
\textit{Stochastic Epidemic Models and
Their Statistical Analysis}, Lect. Notes Stat., Vol. \textbf{151}
(Springer, New York, 2000).
\bibitem{bio} M.S. Samoilov and A.P. Arkin, Nature Biotech. \textbf{24}, 1235 (2006); M. Assaf and B. Meerson, Phys. Rev. Lett. \textbf{100}, 058105 (2008).
\bibitem{kampen} N.G. van Kampen, \textit{Stochastic Processes in
Physics and Chemistry} (North-Holland, Amsterdam, 2001).
\bibitem{gardiner} C.W. Gardiner, \textit{Handbook of Stochastic Methods}
(Springer Verlag, Berlin, 2004).
\bibitem{noexplosion} We will assume that the stochastic population does not exhibit an unlimited growth (escape to infinite population size), see Ref. \cite{MS}, and $n=0$ is the \textit{only} absorbing state.
\bibitem{Assaf} M. Assaf and B. Meerson, Phys. Rev. E \textbf{74},
041115 (2006).
\bibitem{Assaf1} M. Assaf and B. Meerson, Phys. Rev. Lett. \textbf{97}, 200602 (2006).
\bibitem{bender} C.M. Bender and S.A. Orszag, \textit{Advanced Mathematical Methods for Scientists and Engineers} (Springer, New York, 1999).
\bibitem{kubo} R. Kubo, K. Matsuo, and K. Kitahara, J. Stat. Phys. \textbf{9}, 51 (1973).
\bibitem{dykman} M.I. Dykman, E. Mori, J. Ross, and P.M. Hunt, J. Chem. Phys. \textbf{100},
5735 (1994).
\bibitem{Kamenev1} V. Elgart and A. Kamenev, Phys. Rev. E
\textbf{70}, 041106 (2004).
\bibitem{Kamenev2} V. Elgart and A. Kamenev, Phys. Rev. E \textbf{74}, 041101 (2006).
\bibitem{AKM} M. Assaf, A. Kamenev, and B. Meerson, Phys. Rev. E \textbf{78}, 041123 (2008).
\bibitem{kessler} D.A. Kessler and N.M. Shnerb, J. Stat. Phys.
\textbf{127}, 861 (2007).
\bibitem{gaveau} B. Gaveau, M. Moreau, and J. Toth, Lett. Math. Phys. \textbf{37}, 285
(1996).
\bibitem{Doering} C.R. Doering, K.V. Sargsyan, and L.M. Sander, Multiscale Model. and Simul. \textbf{3},
283 (2005).
\bibitem{Assaf2} M. Assaf and B. Meerson, Phys. Rev. E \textbf{75}, 031122
(2007).
\bibitem{turner} J.W. Turner and M. Malek-Mansour, Physica A
\textbf{93}, 517 (1978).
\bibitem{MS} B. Meerson and P.V. Sasorov, Phys. Rev. E \textbf{78}, 060103(R) (2008).
\bibitem{EsK} C. Escudero and A. Kamenev, Phys. Rev. E \textbf{79}, 041149 (2009).
\bibitem{nontrivial} At nontrivial fixed points of the rate equation, $q=q_i=n_i/N>0$, the two real roots of the equation
$H(q,p)=0$ merge at $p=0$.
\bibitem{q0} If $S^{\prime\prime}(0)=0$, then it is more convenient to use Eq.~(\ref{s1fast}).
\bibitem{darroch} J.N. Darroch and E. Seneta, J. Appl. Probab. \textbf{4}, 192 (1967).
\bibitem{physical} We remind the reader that, in order to obtain  Eq.~(\ref{rateexp}), we rescaled the reaction rates and
time by the linear decay rate constant $w_{-1}^{\prime}(0)=\alpha$. To express the MTE in physical units one needs to put the factor $\alpha$ back.
\bibitem{AKM1} M. Assaf, A. Kamenev, and B. Meerson, Phys. Rev. E \textbf{79}, 011127 (2009).
\bibitem{Nasell} I. N{\aa}sell, J. Theor. Biol. \textbf{211}, 11 (2001).
\bibitem{Ovas} O. Ovaskainen, J. Appl. Prob. \textbf{38}, 898 (2001).
\bibitem{diffR0} In Refs.~\cite{Nasell,Ovas} the parameter $R_0$ was defined as $\lambda (N-1)/\mu$. Therefore, their result for the MTE looks slightly
different, but it is actually identical to Eq.~(\ref{tausis}) in the leading and subleading orders of $N$.
\bibitem{dykman80} M.I. Dykman and M.A. Krivoglaz, Physica A \textbf{104}, 480 (1980).
\bibitem{3/2} The distance to bifurcation $\delta_1$ is defined in Refs. \cite{dykman} and \cite{dykman80} differently than in the present work. Their parameter $\delta_1$ is related to our parameter $\delta$ as $\delta_1 = \delta^2$, so the entropy barriers in Refs. \cite{dykman} and \cite{dykman80} scale with $\delta_1$ as $\delta_1^{3/2}$.

\end{thebibliography}
\end{document}